\documentclass[11pt,draft]{article}
 \usepackage{amsfonts}
 \usepackage{amssymb}
 \usepackage{cite}
 \def\bsh{\backslash}
 \def\bdt{\dot \beta}
 \def\adt{\dot \alpha}

 \newfont{\bbbold}{msbm10}
 \def\com{\mbox{\bbbold C}}

 \def\bbF{\mbox{\bbbold F}}

 \def\bbM{\mbox{\bbbold M}}

 \def\bbP{\mbox{\bbbold P}}

 \def\bbT{\mbox{\bbbold T}}

 \def\bbZ{\mbox{\bbbold Z}}

 \def\cO{{\cal O}}

 \newfont{\goth}{eufm10 scaled \magstep1}
 
 \def\gb{\mbox{\goth b}}

 \def\gg{\mbox{\goth g}}
 \def\gh{\mbox{\goth h}}

 \def\gl{\mbox{\goth l}}
 
 \def\gn{\mbox{\goth n}}
 
 \def\gp{\mbox{\goth p}}

 \def\gs{\mbox{\goth s}}
 
 \def\gu{\mbox{\goth u}}

 \def\a{\alpha}
 \def\b{\beta}
 \def\c{\gamma}\def\cdt{\dot\gamma}
 \def\d{\delta}\def\D{\Delta}
 \def\vare{\varepsilon}
 \def\vf{\varphi}

 \def\l{\lambda}
 \def\m{\mu}

 \def\r{\rho}
 \def\s{\sigma}
 \def\t{\tau}
 \def\th{\theta}
 
 \def\be{\begin{equation}}\def\ee{\end{equation}}
 \def\bea{\begin{eqnarray}}\def\eea{\end{eqnarray}}
 \def\ba{\begin{array}}\def\ea{\end{array}}
 \def\x{\xi}
 \def\O{\Omega}
 \def\del{\partial}
 \def\ua{\underline{\alpha}}
 \def\ub{\underline{\phantom{\alpha}}\!\!\!\beta}

 \def\unA{\underline A}
 \def\unB{\underline B}

 \def\xz{\times}

 \def\del{\partial}
 
 \def\bt{\bullet}

 \let\la=\label

 {}

 \def\bd{\begin{document}}
 \def\ed{\end{document}}
 \def\bea{\begin{eqnarray}}\def\barr{\begin{array}}\def\earr{\end{array}}
 \def\eea{\end{eqnarray}}
 \def\ft#1#2{{\textstyle{{\scriptstyle #1}\over {\scriptstyle #2}}}}
 \def\fft#1#2{{#1 \over #2}}
 \newcommand{\eq}[1]{(\ref{#1})}
 \def\eqs#1#2{(\ref{#1}-\ref{#2})}
 \def\det{{\rm det\,}}
 \def\tr{{\rm tr}}\def\Tr{{\rm Tr}}

\begin{document}

 \thispagestyle{empty}

 \hfill{KCL-MTH-01-39 }

 \vspace{20pt}

 \begin{center}
 {\Large{\bf Superfield representations of superconformal groups}}
 \vspace{30pt}

 {P.J.Heslop} \vskip 1cm {Department of Mathematics}
 \vskip 1cm {King's College, London} \vspace{15pt}

 \vspace{60pt}

 {\bf Abstract}

 \end{center}

 Representations of four dimensional superconformal groups
  are constructed as fields on many different superspaces, including
  super Minkowski space, chiral superspace, harmonic superspace and
  analytic superspace. Any unitary irreducible representation can be
  given as a field on any one of these spaces if we include fields which transform under supergroups. In particular, on
  analytic superspaces, the fields are unconstrained. One can obtain
  all representations of the $N=4$ complex superconformal group $PSL(4|4)$ with integer dilation
  weight from copies of the Maxwell
  multiplet on $(4,2,2)$ analytic superspace. This construction is
  compared with the oscillator construction and it is shown that
  there is a natural correspondence between
  the oscillator construction of superconformal representations
  and
  those carried by superfields on analytic superspace.

 {\vfill\leftline{}\vfill \vskip  10pt

 \baselineskip=15pt \pagebreak \setcounter{page}{1}


\section{Introduction}

Recently there has been renewed interest in the study of unitary
irreducible representations of superconformal groups in the light
of the AdS/CFT correspondence \cite{Maldacena:1998re}. This correspondence
relates M theory or string theory on $AdS \times S$ backgrounds to
superconformal field theories which can be thought of as living on
the boundary of the $AdS$ space. The bulk theory and the boundary
theory both have the same symmetry group, and the operators in
both theories fall into representations of the superconformal
group.

There are different methods of constructing these representations
\cite{Flato:1984te,Dobrev:1985qv,Dobrev:1987qz,Binegar:1986ib,Morel:1985if,
 Bars:1983ep,Gunaydin:1985fk,Gunaydin:1998sw,Gunaydin:1998jc,Gunaydin:1999jb}.
 In this paper we show how to construct any
unitary irreducible representation of the four dimensional
superconformal groups $SU(2,2|N)$ explicitly, acting on
superfields on a number of different superspaces. These spaces
include super Minkowski space, chiral superspaces, harmonic
superspaces and analytic superspaces. Until
recently~\cite{Heslop:2000mr,Heslop:2001dr,Heslop:2001gp} only
superfields which transform trivially under semisimple supergroups
have been considered. It is by allowing superfields to carry
non-trivial representations of supergroups (i.e. by allowing
superfields with superindices) that we are able to obtain all
unitary irreducible representations as superfields. In particular,
we show that all unitary irreducible representations are carried
by unconstrained fields on analytic superspaces.

The techniques here may be thought of as the natural application
of twistor techniques to the supersymmetric case. However things
are much simpler in this case as we need not consider higher
cohomology. Indeed it is only on super (ambi-) twistor spaces that
one needs higher cohomology to describe superconformal
representations. For all other superspaces unitary irreducible
representations are carried by superfields.

A number of papers have been written in which superconformal field
theories are studied on analytic superspaces
\cite{Howe:1996rm,Howe:1998zi,Eden:1999gh,Eden:1999kw,Eden:2000qp,Eden:1998hh,
 Eden:2000mv}. The
advantage of these spaces is that the fields are automatically
irreducible and thus the full power of superconformal invariance
can be exploited more easily. By allowing fields which transform
non-trivially under supergroups we are able to give all unitary
irreducible representations as unconstrained fields on analytic
superspace. Therefore representations not previously considered in
the analysis can be included, and in principle the whole theory
could be given in analytic superspace
(see~\cite{Heslop:2001dr,Heslop:2001gp} for recent work using
these techniques).

In section~\ref{fields} we discuss on-shell massless multiplets
and their manifestation as constrained fields on super Minkowski
space. We review harmonic superspaces, first introduced by GIKOS
\cite{Galperin:1984av,Galperin:1984ec,Galperin:1985bu,
Galperin:1987id,Galperin:1985zv},
and how to obtain all series B and C representations as analytic
fields on these, which are invariant under supergroups
\cite{Heslop:2000af}. We then have a brief look at representations
on $N=2$ analytic superspace and $(4,2,2)$ analytic superspace
which transform under supergroups.

In section~\ref{supergroups} we review various aspects of
supergroup representation theory using the method of parabolic
induction. We consider super Dynkin diagrams (or Kac-Dynkin
diagrams), which will have crosses through to indicate the space,
and numbers above the nodes to indicate the representation. We review
the unitary bounds and give these as bounds on the allowed Dynkin labels.

In section~\ref{sect:red}
we discuss the finding of irreducible representations firstly in the
context of Verma modules, and then translate this to Minkowski space,
giving the constraints fields must satisfy in order to carry
irreducible representations. We then apply these techniques to the
supersymmetric case and
give the constraints that superfields on super Minkowski space
must satisfy in order to carry unitary irreducible
representations. Finally we show that these correspond to unconstrained
superfields on analytic superspace.

In section~\ref{sindices} we look at finite dimensional
representations of supergroups using superindices and super Young
tableaux. We show how to construct any representation on any
allowed space as a tensor (or quasi-tensor) superfield. We
consider examples in $N=2$ analytic superspace and show how to
construct all representations (with quantised dilation weight) in
$(4,2,2)$ analytic superspace from the basic Maxwell multiplet.

In section~\ref{oscillator} we look at the oscillator method of
obtaining representations. Firstly we look at representations of
the conformal group and show the relationship between oscillator
representations and representations as conformal fields on
Minkowski space. If we apply this method for superconformal
representations one is naturally lead to representations carried
by superfields on analytic superspace.

There is an appendix in which we review various aspects of
conformal representations as fields on Minkowski space and the
irreducibility conditions they satisfy. The results of this paper
come from applying these techniques in the superspace context.


\section{Superconformal fields}\label{fields}


\subsection{Massless multiplets on super Minkowski space} \label{massless}

An important class of operators consists of those which undergo
shortening since they are protected from renormalisation. On super
Minkowski space the shortening conditions occur as differential
operators acting on the superfields. For example, on-shell
massless supermultiplets can be given as constrained fields on
super Minkowski space
\cite{Siegel:1981bp,Howe:1981qj,Howe:1981xy}. There are three
classes of massless supermultiplets.

Firstly there are supermultiplets with maximal helicity $s$, with
$\lceil\frac N2\rceil \leq 2s < N$, ($\lceil\frac N2\rceil$ denotes
the nearest
integer greater than or equal to $\frac N2$). These are described in
(real) super Minkowski
space, $M$, by superfields $W$ which have $p=2s$ totally
antisymmetric internal indices and which satisfy
\begin{eqnarray}
\bar D_{\adt}^i W_{j_1 \ldots j_{p}} & = &
\frac{p(-1)^{p-1}}{N-p+1} \delta_{[j_1}^i \bar D_{\adt}^k W_{j_2 \ldots
j_{p}] k} \nonumber \\ D_{\alpha i}W_{j_1 \ldots j_{p}} & = &
D_{\alpha [ i} W_{j_1 \ldots j_{p}]} \label{DW}
\end{eqnarray}
For each such superfield there is a conjugate superfield $\tilde
W_{i_1\ldots i_{N-p}}$ defined by \be \tilde W_{i_1\ldots
i_{N-p}}={1\over p!}\vare_{i_1\ldots i_{N-p}j_{N-p+1}\ldots j_N}
\bar W^{j_{N-p+1}\ldots j_N} \ee satisfying similar constraints.
When $s=\frac{1}{4}N \in \bbZ^+$, the multiplet is self-conjugate:
\begin{equation}
\bar{W}^{i_1 \ldots i_{p}} = \frac{1}{p!} \varepsilon^{i_1 \ldots
i_{p} j_1 \ldots j_{p}} W_{j_1 \ldots j_{p}}.
\end{equation}
We therefore have scalar superfields $W_{i_1\ldots i_p}$
antisymmetric on $p$ internal indices with $p$ ranging from 1 to
$N-1$ all of which obey the constraints \eq{DW}. The second case
extends this to $p=0$ and $p=N$ where we have $s={N \over 2}$ and
\eq{DW} implies that the field is chiral or anti-chiral
respectively. These must satisfy a further second order constraint
in order to be irreducible. So the chiral field $W$ satisfies
 \be
 \bar{D}_{\adt}^i W =0, \qquad D_{\a i} D^{\a}_j W=0.
 \la{chiraleom}
 \ee
 and the antichiral field $\bar{W}$ satisfies
 \be D_{\a i} \bar{W} =0, \qquad \bar{D}_{\adt}^i D^{\adt j}
 \bar{W}=0. \ee
 The third case consists of massless multiplets with
 maximal helicity $s=J_1 + {N \over
  2}$ which correspond to chiral
 superfields with spacetime indices $W_{\a_1 \cdots \a_{2J_1}}$. These
  must satisfy another constraint in addition to the chiral constraint in
  order to form an irreducible representation of the superconformal
group:
 \be
 \bar{D}_{\adt}^i W_{\a_1 \cdots \a_{2J_1}}=0 \qquad D^{\a}_iW_{\a \a_2
  \cdots \a_{2J_1}}=0.\label{chiralind}
 \ee Conjugate massless multiplets are defined similarly as
 antichiral fields $W_{\adt_1 \cdots \adt_{2J_2}}$ satisfying the
  complex conjugate constraints
\be
D_{\a i} W_{\adt \adt_2 \cdots \adt_{2J_2}}=0 \qquad \bar D^{\adt
i} W_{\adt \adt_2 \cdots \adt_{2J_2}}=0.
\ee

\subsection{Harmonic superspaces}

 Harmonic superspaces are usually taken to be of the form $M \xz F$
 where $M$ is super Minkowski space and $F=P\bsh G$ is a coset
 space of the internal symmetry
 group $G$ with isotropy group $P$
 \cite{Galperin:1984av,Galperin:1984ec,Galperin:1985bu,Galperin:1987id,
 Howe:1995md,Hartwell:1995rp}.
 The harmonic superspace with
 $P=S(U(p) \xz U(N-p-q) \xz U(q))$ is known as $(N,p,q)$ harmonic
 superspace which we will use here. Generalised $(N,p,q)$ harmonic
 superspaces can be defined similarly, but with $U(N-p-q)$ replaced
 with some subgroup of $U(N-p-q)$.

 In the GIKOS formalism \cite{Galperin:1984av} one instead works with
 fields defined
 on $M\xz G$
 which are equivariant with respect to $P$. That is, one has fields
 $A(z,u),\ z\in M, u\in G$ which take their values in some
 representation space of $P$ and which satisfy $A(z,hu)= R(h)A(z,u)$,
 $R$ being the representation of $P$ in question. In
 index notation we write $u_I{}^i$ where the upper case index
 indicates a representation of $H$ and the lower case index the
 defining representation of $G$, and where both indices run from $1$ to $N$.
 We can use $u$ to convert $G$ indices to $H$ indices. For
 $(N,p,q)$ harmonic superspaces we split
 the index $I$ as follows under $H$: $I=(r,r'',r')$, where
 $r=1,\ldots p,\ r''=p+1,\ldots N-q,\ r'=N-q+1,\ldots N$,
 and where $(p+q)\leq N$, then we can have superfields $A$ which obey
 the constraints
 \be
 D_{\a r}A=\bar D_{\adt}^{r'}A=0
\la{pq}
 \ee
 where
 \be D_{\a r}:=u_r{}^i D_{\a i};\qquad \bar D_{\adt}^{r'}=\bar
 D_{\adt}^i(u^{-1})_i{}^{r'}.
 \ee
 Constraints such as these are called Grassmann analyticity
 (G-analyticity) conditions following \cite{Galperin:1984ec}. The
 derivatives on
 $G$ can be taken to be the
 right-invariant vector fields, $D_I{}^J,\ D_I{}^I=0$. They act as
 follows:
 \be
 D_I{}^J u_K{}^k=\d_K{}^J u_I{}^k-{1\over N}\d_I{}^J u_K{}^k
 \ee
 and they obey the reality condition
 \be
 \bar D^J{}_I=-D_I{}^J.
 \ee
 They divide into three sets; the set
 $\{D_{r}{}^{s'},D_{r}{}^{s''},D_{r''}{}^{s'}\}$, corresponding to the
 $\bar\del$ operator on the complex coset space (i.e. holomorphic
 functions are annihilated by $\bar\del$), the complex conjugate set
 $\{D_{r'}{}^{s},D_{r''}{}^{s},D_{r'}{}^{s''}\}$ which corresponds to
 the $\del$ operator, and the set
 $\{D_{r}{}^{s},D_{r'}{}^{s'},D_{r''}{}^{s''}\}$ which corresponds to
 the isotropy algebra. As well as imposing G-analyticity we can impose
 harmonic analyticity (H-analyticity) by demanding that a superfield
 be annihilated by the first of these sets of derivatives. Indeed,
 since the algebra of all of these operators and the spinorial
 derivatives defining G-analyticity closes, it is consistent to impose
 both G-analyticity and H-analyticity on the same superfield. Such
 superfields will be referred to as CR-analytic, or simply analytic.

 Many of the massless multiplet fields defined above are described
 naturally as analytic fields on harmonic superspace. The
 superfields $W_{i_i\ldots i_p}$, satisfying
 \eq{DW}, can be naturally described as
CR-analytic fields on $(N,p,N-p)$ harmonic superspaces
 \cite{Hartwell:1995rp}. Define the scalar superfield
 \be W=\frac{1}{p!}\vare^{r_1 \cdots
 r_p}u_{r_1}{}^{i_1} \cdots u_{r_p}{}^{i_p}W_{i_1
 \cdots i_p}\label{W}
 \ee
 on $(N,p,q)$ harmonic superspace. Then the conditions \eq{DW} become
 G-analyticity constraints on $W$
 \be \ba{rcl} D_{\a r} W^{(p)} &= & 0\\ \bar D_{\adt}^{r'}W^{(p)} &=&0
 \la{gan},
 \ea \ee
 and the H-analyticity constraint,
 \be
 D_r{}^{s'}W^{(p)}=0 \la{han}
 \ee
 and we see that $W$ is equivariant with respect to the isotropy group
 $S(U(p) \xz U(N-p)$. Conversely, an equivariant field with the same
 charge (under the $U(1)$ subgroup of the isotropy group) as $W$ and
 satisfying the above analyticity constraints is equivalent to a
 superfield on Minkowski superspace satisfying \eq{DW}. We can apply this
 construction to the entire family of scalar superfields with $p$
 varying from $0$ to $N$, bearing in mind that the chiral fields
 satisfy the additional constraint \eq{chiraleom}.

 Whilst $(N,p,N-p)$ harmonic superspace is perhaps the most natural
 harmonic superspace on which to represent these massless multiplets,
 there are many other ways of representing them that are ``less efficient''
 in that the superspaces have more odd coordinates \cite{Heslop:2000af}. For
 example, any of the above massless multiplets, $W^{(p)}$, (except $p= 0, N$)
 can be realised with $(1,1)$ G-analyticity on an internal space
 $F=U(1)^{N-1}\bsh SU(N)$. We define
 \be
 W_{1 \dots p} = u_1{}^{i_1} \dots u_p{}^{i_p} W_{i_1 \dots i_p}
 \ee
 where $W_{i_1 \dots i_p}$ is the superfield on super Minkowski space.
 Equation \eq{DW} is equivalent to the G-analyticity
 conditions
 \bea
 D_{\a 1} W_{1 \dots p} &=& 0\\
 \bar D_{\adt}^N W_{1 \dots p} &=&0,
 \eea
 and the H-analyticity conditions
 \be
 D_I{}^J W = 0 \mbox{ for } I < J.
 \ee

 Unitary representations of the superconformal group fall into
 three series which have been called A, B and C (see
 section~\ref{ubounds}). Series B and C correspond
 to short multiplets.
 There is a very simple way of generating all possible series B
 and C representations from the set of available massless supermultiplets
 for a given $N$ using this harmonic superspace
 \cite{Heslop:2000af,Ferrara:2000eb}.
 Firstly, for series C, if $(a_1, \dots,
 a_{N-1})$ are the internal group Dynkin labels of the
 representation we are interested in, simply form the product
 \be A=\prod_{k=1}^{N-1}\left(W_{12\ldots k}\right)^{a_k}. \la{prod}
 \ee
 The dilation weight is equal to the sum of the Dynkin labels,
 $m_1=\sum_{k=1}^{N-1}a_k$, since each underlying field has
 dilation weight 1. Furthermore the field $W_{12\ldots k}$ has
 R-charge $ {2k} \over N -1$ and so the field $A$ will have
 R-charge $2m \over N -m_1$ where $m=\sum_{k=1}^{N-1}k a_k$.

 For series B representations we work on $(N,0,1)$
 superspace, again with the internal
 manifold $F=U(1)^{N-1}\bsh SU(N)$. We use the same
 family of superfields as before augmented with the chiral fields from
 the previous section, $W$, $W_{\a_1, \dots, \a_{2J_1}}$. On this
 harmonic superspace these satisfy both analytic and non-analytic
 constraints (from~(\ref{chiraleom},\ref{chiralind})):
 \be \ba{rclcrcl}
 \bar D_{\adt}^N W&=&0& \qquad &D_{\a 1}D^{\a}_1 W &=&0\\
 \bar D_{\adt}^N W_{\a_1 \dots \a_{2J_1}}&=&0 &\qquad & D^{\a}_1
 W_{\a \a_2 \dots \a_{2J_1}}&=&0
 \ea. \ee
 We can then form the product superfield~\cite{Heslop:2000af,Ferrara:2000eb}
 \be A_{\a_1 \dots \a_{2J_1}}= W_{\a_1 \dots \a_{2J_1}}
 W^b\prod_{k=1}^{N-1}\left(W_{12\ldots k}\right)^{a_k}
 \ee
 where $b$ is a positive number (we allow non-integer powers to obtain
 representations with non-integer dilation weight). The $L$ weight of
 this field is
  $L=m_1+b+J_1$ and the R-symmetry weight is  $R=2m/N-(m_1+b+J_1+1)$
 and therefore
 $L+R=2m/N$ as required (see~(\ref{abc})). In the case $b=0$ these
 fields will
 satisfy the additional constraints  $D^{\a}_1
 A_{\a \a_2 \dots \a_{2J_1}}=0$ if $J_1$ is non-zero, or $D_{\a
 1}D^{\a}_1 A=0$ if $J_1$ is zero. These representations saturate the
 bounds of~(\ref{abc}).

 We can also produce the series A fields if we also include antichiral
 fields $\bar{W}$ and $\bar{W}_{\adt_1, \dots, \adt_{J_2}}$. We form
 the field~\cite{Ferrara:2000eb}
 \be
 W_{\a_1 \dots \a_{2J_1}}\bar{W}_{\adt_1 \dots \adt_{J_2}}
 W^b \bar{W}^c \prod_{k=1}^{N-1}\left(W_{12\ldots k}\right)^{a_k}
 \ee
 where $b,c\geq0$. This field is not an analytic superfield and in
 general is unconstrained although in the case $b=0$ or $c=0$
 we obtain further constraints.

 \subsection{All representations as analytic fields}\la{sec:an}

 In this paper we give a systematic study of
 unitary superconformal representations as superfields on
 different superspaces with the aid of super Dynkin diagrams.
 In fact all
 unitary irreducible representations of the superconformal group can be given
 as superfields on many different superspaces, and in particular on analytic
 superspaces all fields carry irreducible representations.
 We complexify spacetime and the superconformal group to $SL(4|N)$. Then
 analytic superspaces can be exhibited as coset spaces even though
 they are not coset spaces of the real superconformal group
 $SU(2,2|N)$. On complex $(N,p,q)$ superspace fields transform under
 the group $S(GL(2|p) \xz GL(N-p-q) \xz GL(2|q))$. Only
 fields which are
 invariant under the supergroups $SL(2|p)$ and $SL(2|q)$ have been
 considered until recently \cite{Heslop:2000mr,Heslop:2001dr,Heslop:2001gp}.
 If we relax this and
 allow fields that transform
 under supergroups then all representations can be given as fields
 on any superflag space (except super (ambi-) twistor spaces for technical reasons we will
 explain later.)

 For example,  consider $N=2$ analytic superspace. Fields on this
 space transform under $S(GL(2|1) \xz GL(2|1))\bsh
 SL(4|2)$. The coordinates of analytic superspace are
 \be
 X^{AA'}=\left( \ba{cc} x^{\a \adt} & \l^{\a} \\
                        \pi^{\adt}  &  y  \ea \right)
 \ee
 where $x$ are the usual (complex) spacetime coordinates, $\l,\pi$ are odd
 coordinates, and $y$ is the coordinate of the internal manifold $\com
 P^1$.
 We have already seen (in the real case) that the hypermultiplet can
 be exhibited as an analytic
 field. The hypermultiplet is an $N=2$ massless
 multiplet with highest spin $s={1 \over 2}$ and so it is given on
 super
 Minkowski space by a field $W_i$
 satisfying equations~\eq{DW} and on harmonic superspace by a field
 $W$ satisfying the analytic constraints (\ref{gan},\ref{han}). In the
 complex setting this is an
 unconstrained field on analytic superspace invariant under both $SL(2|1)$
 supergroups but transforming with a $\com$-weight.

 A more unusual multiplet to consider as an analytic field is the $N=2$
 Maxwell multiplet. This is normally given as a chiral
 field on super Minkowski space, satisfying the additional
 constraint~\eq{chiraleom}. On analytic superspace it is
 given as a holomorphic field $W_A$ transforming contravariantly under
 the first $SL(2|1)$ subgroup, and satisfying no constraints. The
 conjugate multiplet is given similarly by a field $W_{A'}$
 transforming under the second $SL(2|1)$ subgroup.

 Another example is
 the $N=2$ superconformal stress-energy tensor, which on super
 Minkowski space is a scalar superfield satisfying a second order
 constraint. On analytic superspace it is given by a field $T_{A'A}$
 transforming under both $SL(2|2)$ supergroups. Once more this
 irreducible representation is given as an unconstrained field on analytic
 superspace. This representation can also be realised by multiplying
 massless multiplets together in two
 different ways on analytic superspace: firstly by multiplying
 a Maxwell field and its conjugate together
 \be
 T_{A'A}= W_{A'} W_{A}\la{WW}
 \ee
 and secondly by multiplying two hypermultiplet fields together
with a derivative:
 \be
 T_{A'A}=W_1\partial_{A'A}W_2 -W_2\partial_{A'A}W_1 \la{WdW}.
 \ee

 In the case $N=4$ consider $(4,2,2)$ analytic superspace. The
 coordinates of $(4,2,2)$ analytic superspace are given as
 \be
 X^{AA'}=\left( \ba{cc} x^{\a \adt} & \l^{\a a'} \\
                        \pi^{a\adt}  &  y^{aa'}  \ea
 \right) \label{N=4}
 \ee
 where again $x$ are spacetime coordinates, $\l,\pi$ are odd
 coordinates and $y$ are coordinates for the internal manifold.
 The group $SL(4|4)$ is not semisimple and thus one normally divides
 out the centre to leave $PSL(4|4)$. For representations this simply
 means only those with vanishing R-charge are allowed. For fields on analytic superspace the condition $R=0$
 turns out to be the condition that the number of primed indices must
 equal the number of unprimed indices. The only massless multiplet
 with vanishing R-charge
 is the $N=4$ Maxwell multiplet. Given on super Minkowski space
 as the field $W_{ij}$ it becomes the scalar field $W$ on analytic
 superspace, defined as in \eq{W}.
 As we will argue later one can in fact produce all representations of
 the supergroup $PSL(4|4)$ using copies of the Maxwell multiplet,
 together with derivatives. For example the field
 \be
 W_{A'A}=W^{(1)}\del_{A'A}W^{(2)} - W^{(2)}\del_{A'A}W^{(1)}\label{comb}
 \ee
 where $W^{(i)}$ are different copies of the Maxwell multiplet on
 $(4,2,2)$ analytic superspace, corresponds to the series C representation with
 internal Dynkin labels
 $[1,0,1]$ and dilation weight $L=2$.


\section{Representations of supergroups}\label{supergroups}


We wish to apply the techniques outlined in the appendix for Lie
groups to simple Lie supergroups, concentrating in particular on
the complex group $SL(r|s)$ and sometimes specifically the complex
superconformal group $SL(4|N)$. Many concepts can be
carried over from the purely bosonic case with some important
differences which we will point out when they occur. See
\cite{Cornwell} for a review of super Lie groups and their
representations.

\subsection{Roots and parabolic subalgebras of $SL(r|s)$}

We set the Cartan subalgebra $\gh$ of $\gs \gl(r|s)$ to be the set
of supertraceless diagonal matrices. The roots are given by
$e_{ij}\in \gh^*$ with $1 \leq i \neq j \leq r+s$ where \be
e_{ij}(h) = h_i -h_j \qquad \gh \ni h = \mbox{diag}( h_1, h_2,
\dots, h_r | h_{r+1}, h_{r+2}, \dots, h_{r+s})
 \ee
and where the $h_i$ are any Grassmann even numbers, subject to the
constraint
 \be
 \sum_{k=1}^r h_k -\sum_{k=r+1}^{r+s} h_k = 0.
 \ee The
subspaces of $\gs \gl(r|s)$, $\gg_{e_{ij}}$ corresponding to these
roots are spanned by the matrices $\hat{e}_{ji}$ which have zeros
everywhere except in
 the $ji$ component where there is a 1. A root is said to be even
 or odd according to
whether the non zero component of $\gg_{e_{ij}}$ is even or odd.
We now choose a system of simple roots, for example for $SL(4|N)$
the set of roots \be S_D=\{e_{i(i+1)}:i=1 \dots 3+N\}. \ee This
particular choice of roots is called the distinguished system of
simple roots as it has only one odd root $e_{45}$. However unlike
in the purely bosonic case, where all choices of simple roots are
equivalent up to conjugation, we can now have inequivalent systems
of simple roots. For example, another choice for a system of
simple roots for $SL(4|N)$, which will turn out to be the one we
will use, is \be S=\{e_{12}, e_{25}, e_{67}, \dots ,
e_{(3+N)(4+N)}, e_{(4+N)3}, e_{34}\}. \label{S}\ee This has two
odd roots $e_{25}$ and $e_{(4+N)3}$. The easiest way to see that
these are inequivalent is by constructing the respective standard
Borel subalgebras from them.

Given a choice of simple roots we define the set of positive roots,
$\D^+$, in the usual  way, as in~(\ref{D+})
and then the standard Borel subalgebra $\gb$ is defined as
in~(\ref{borel},\ref{nil}). We see that the standard Borel subalgebra
for $SL(4|N)$ constructed with the choice of simple roots $S_D$,
is simply the set of lower triangular matrices \be
\gb_{S_D}=\left\{
\left(\ba{cccc|ccc} \bt &&&&&&\\ \bt&\bt&&&&&\\ \bt &\bt & \bt &&&&\\
\bt &\bt & \bt &\bt &&&\\
\hline \bt &\bt & \bt &\bt &\bt &&\\\bt &\bt & \bt &\bt & \cdot
&\cdot &\\ \bt &\bt & \bt &\bt & \bt & \cdot & \bt \ea
\right)\right\} \ee whereas the Borel subalgebra constructed using
$S$ is given by matrices with the following form: \be
\gb_{S_D}=\left\{ \left(\ba{cccc|ccc} \bt &&&&&&\\ \bt&\bt&&&&&\\
\bt & \bt &\bt&& \bt &\cdot& \bt\\ \bt &\bt &\bt &\bt&\bt
&\cdot&\bt\\\hline
\bt &\bt &&&\bt &&\\ \cdot & \cdot &&&\cdot &\cdot &\\
\bt &\bt &&&\bt &\cdot&\bt\\ \ea \right)\right\}. \ee However, if
we make a simple change of basis for $\com^{4|N}$ on
 which the
 $SL(4|N)$ matrix acts, we can change an element in the Lie
algebra $\gs \gl(4|N)$ as follows: \be
          g= \left( \ba{cc|c} \raisebox{-1.5ex}{\makebox[1em][l]{\hspace{.4em} 4}}&\phantom{4}& \\ \phantom{4}&\phantom{4}&\\ \hline &&N
                          \ea \right)
  \rightarrow
  \left( \ba{c|c|c}  2&& \\ \hline &N& \\ \hline &&2 \ea     \right). \label{eq:g}
  \ee
In this new basis the Borel subalgebra becomes lower triangular
again \be \gb_{S}=\left\{ \left(\ba{cc|ccc|cc} \bt &&&&&&\\
\bt&\bt&&&&&\\ \hline \bt&\bt&\bt&&&&\\
\cdot&\cdot&\cdot&\cdot&&&\\
\bt&\bt&\bt&\cdot&\bt&&\\ \hline
\bt&\bt&\bt&\cdot&\bt&\bt&\\
\bt&\bt&\bt&\cdot&\bt&\bt&\bt\\ \ea \right) \right\}. \ee For the
supergroup $SL(r|s)$, for any choice of simple roots, we can
always choose a basis for $\com^{r|s}$ on which the group acts
such that the Borel subgroup is the set of lower triangular
matrices. We see from our example that this change of basis can
not be performed using an $SL(r|s)$ matrix as it swaps bosonic and
fermionic components around. So $\gb_S$ is not equivalent to
$\gb_{S_D}$ under conjugation (i.e. $\gb_S \neq g \gb_{S_D}
g^{-1}$ for any $g \in G$) and the choices of simple roots are
inequivalent.

 The Killing form for a superalgebra $\gg$ is defined as
 \be
 (u,v)= \mbox{ str ad}(u)\mbox{ ad}(v)
 \ee
 for $u, v \in \gg$, where ad denotes the adjoint representation.
 This simplifies in the case $SL(r|s)$ to
 \be (u,v) = 2(r-s)\mbox{ str }uv \qquad r
 \neq s.
 \ee
In the case $r=s$ this vanishes but a convenient non-degenerate
 bilinear form that can be used in
 its place is provided by~\cite{Cornwell}
\be
(u,v) = \mbox{ str }uv
\ee
and we shall mean this, when $r=s$, when we refer to the Killing form.
We use this Killing form to identify $\gh$ with $\gh^*$ just as in
the purely bosonic case~(\ref{ha}). However $(\a,\a)$ is no longer
necessarily positive for any root $\a$. Indeed, if $\a$ is an odd
root it is possible that $(\a,\a)=0$. We define the co-root
$\a^{\vee}$ of a simple root $\a$ as \be \a^{\vee} = \left\{
\ba{cc}
2\a/(\a,\a) & \mbox{if }(\a,\a) \neq 0\\
\a/(\a,\b)  & \mbox{if }(\a,\a) =0\\ \ea \right. \ee where we
choose a simple root $\b \in S$ such that $(\a,\b)\neq0$. The
Cartan matrix is defined just as in~(\ref{cartan}). The Dynkin
diagram is defined from the Cartan matrix in exactly the same way
as in the bosonic case, except we now distinguish between even and
odd simple roots by assigning black and white nodes to them
respectively.
The Dynkin diagram for $\gg$ is not unique and depends on the
choice of simple roots, inequivalent systems of simple roots
giving rise to different Dynkin diagrams.

The Dynkin diagram for $SL(4|N)$ obtained from the distinguished
system of simple roots $S_D$ is \be
\begin{picture}(300,20)(-10,-10)
\put(0,0){\makebox[0pt][l]{$\bt\hspace{1.5em}\bt\hspace{1.5em}
\bt\hspace{1.5em}\ominus\hspace{1.5em}
    \underbrace{\bt \hspace{1.5em}
\bt\hspace{1.5em}\cdots
\hspace{1.5em}\bt\hspace{1.5em}\bt}_{N-1}$}
\rule[.5ex]{11.4em}{.1ex} $\hspace{4.3em}$ \rule[.5ex]{2.2em}{.1ex}
 }
\end {picture}
\ee Here the three nodes on the left correspond to the $\gs
\gl(4)$ subgroup of $\gs \gl(4|N)$, the $N-1$ nodes on the right
correspond to the $\gs \gl(N)$ subgroup of $\gs \gl(4|N)$, and the
white node corresponds to the odd root. The Dynkin diagram
obtained using the system of simple roots $S$ is
 \be
\begin{picture}(300,20)(-10,-10)
\put(0,0){\makebox[0pt][l]{$\bt\hspace{1.5em}\ominus\hspace{1.5em}
    \underbrace{\bt \hspace{1.5em}
\bt\hspace{1.5em}\cdots
\hspace{1.5em}\bt\hspace{1.5em}\bt}_{N-1}\hspace{1.5em}\ominus\hspace{1.5em}
\bt$} \rule[.5ex]{6.85em}{.1ex} $\hspace{4.5em}$
\rule[.5ex]{6.6em}{.1ex}
 }
\end {picture}\label{dynk}
\ee

Parabolic subalgebras are ones which contain the Borel subalgebra,
and they can be constructed in the same way as in the bosonic
case, by specifying a subset $S_{\gp}$ of the simple roots $S$ and
finding $\gl$ and $\gn$ (see~(\ref{D}-\ref{p}). For $\gs
\gl(r|s)$, once we have obtained a basis for $\com^{r|s}$ with
respect to which the Borel subalgebra consists of lower triangular
matrices, the parabolic subalgebras are simply the sets of block
lower triangular matrices of a particular shape, and the
corresponding Levi subalgebras are the block diagonal parts. They
can be represented on the Dynkin diagram for $\gs \gl (r|s)$ by
crossing out the nodes corresponding to simple roots in $S\bsh
S_{\gp}$.

 All compactified complexified superspaces that have been used
 in the study of four
 dimensional flat space supersymmetric field theories are of the
 form $P\bsh SL(4|N)$ where $P$ is a parabolic subgroup represented by
 putting crosses on the Dynkin diagram~(\ref{dynk}).

 For example, compactified complexified super Minkowski space
 $\tilde{\bbM}$ has the form
 $P\bsh SL(4|N)$ where
 $P$ consists of matrices of the form
 \be
 \left( \ba{cc|ccc|cc}
                         \bt  &  \bt  &&&&&\\
                         \bt  &  \bt  &&&&&\\
\hline                   \bt  &  \bt  & \bt  &. & \bt  &&\\
                        . &.  &. &. &. &&\\
                         \bt  &  \bt  & \bt  &. & \bt  &&\\
\hline                   \bt  & \bt  & \bt  &.  & \bt  & \bt  & \bt \\
                         \bt  & \bt  & \bt  &.  & \bt  & \bt  & \bt
                        \ea \right)\la{smink}
\ee with unit superdeterminant. The corresponding Levi subgroup $L$ is
the set of block diagonal matrices of the form
 \be
 \left( \ba{cc|ccc|cc}
                         \bt  &  \bt  &&&&&\\
                         \bt  &  \bt  &&&&&\\
\hline                     &    & \bt  &. & \bt  &&\\
                        &  &. &. &. &&\\
                           &    & \bt  &. & \bt  &&\\
\hline                   &  &  &  &   & \bt  & \bt \\
                           &   &   &  &   & \bt  & \bt
                        \ea \right)
\ee with unit superdeterminant.
 This parabolic subalgebra can be constructed using
the choice of simple roots $S$ in~(\ref{S}) and with
$S_{\gp}=\{e_{25},e_{(4+N)3}\}$ consisting of the two odd roots.
It has corresponding Dynkin diagram \be
\begin{picture}(300,0)(-5,0)
\put(0,0){\makebox[0pt][l]{$\bt\hspace{1.5em}\otimes\hspace{1.5em}\bt\hspace{1.5em}
\bt\hspace{1.5em}\cdots
\hspace{1.5em}\bt\hspace{1.5em}\bt\hspace{1.5em}\otimes
\hspace{1.5em} \bt$} \rule[.5ex]{6.7em}{.1ex} $\hspace{4.5em}$
\rule[.5ex]{6.9em}{.1ex}
 }
\end {picture}
\ee where the two odd roots have crosses through them. Note that
it is very simple to read off the form of the Levi subalgebra
$\gl$ from the Dynkin diagram. The two crossed through nodes
represent $\com$-charges and the remaining nodes give us the form
of the semisimple part $\gl_S$. Here we obtain
$\gl=\gs\gl(2)\oplus\gs\gl(N)\oplus\gs\gl(2)\oplus\com^2$.

 We can identify non-compact complex super Minkowski space $\bbM$ as an open
 set in compact complex super Minkowski space. The standard coset
 representative is
 \be
  M\ni z\mapsto s(z)=\left( \ba{c|c|c}
  1_2&\th &x  \\
 \hline
 0&1_N &\vf\\
  \hline
   0&0 &1_2\\
  \ea\right).
 \ee
 Here we see the usual coordinates for complex super Minkowski space,
 with $\vf$ denoting the $N$ dotted two-component spinorial coordinates
 which become
 the complex conjugates of the $\th$'s in the real case. We
 can think of the coordinates for a space as fitting into the blank
 spaces of the isotropy group as illustrated in the above case.

 If $P_1 \subset P_2$ are two parabolic subgroups of a group $G$, then
 the coset
 space $P_1\bsh G$ is a fibre bundle over $P_2\bsh G$ with typical
 fibre $P_1 \bsh P_2$
 \begin{equation}
 P_1  \bsh G \longrightarrow P_2  \bsh G .\la{fib}
 \end{equation}
 Non-compact superspaces are open subsets of the coset spaces $P\bsh
 SL(4|N)$ which are obtained by considering fibrations (sometimes
 double fibrations) starting with non-compact super Minkowski space.

For example, left chiral superspace $\tilde{\bbM}_L$ has isotropy
group of the form
 \be
 \left( \ba{cc|ccc|cc}
                         \bt  &  \bt  &&&&&\\
                         \bt  &  \bt  &&&&&\\
\hline                   \bt  &  \bt  & \bt  &. & \bt  &\bt&\bt\\
                        . &.  &. &. &. &.&.\\
                         \bt  &  \bt  & \bt  &. & \bt  &\bt&\bt\\
\hline                   \bt  & \bt  & \bt  &.  & \bt  & \bt  & \bt \\
                         \bt  & \bt  & \bt  &.  & \bt  & \bt  & \bt
                        \ea \right)
\ee
 and has coordinates $(x,\th)$. The corresponding Dynkin diagram is
 \be
 \begin{picture}(300,0)(-5,0)
\put(0,0){\makebox[0pt][l]{$\bt\hspace{1.5em}\otimes\hspace{1.5em}\bt
    \hspace{1.5em}
\bt\hspace{1.5em}\cdots
\hspace{1.5em}\bt\hspace{1.5em}\bt\hspace{1.5em}\ominus
\hspace{1.5em} \bt$} \rule[.5ex]{6.7em}{.1ex} $\hspace{4.5em}$
\rule[.5ex]{7.1em}{.1ex}
 }
\end {picture} \label{chiral}
\ee
 There is a fibration between compactified super Minkowski space and chiral
 superspace of the form \eq{fib}
 \begin{equation}
 \tilde{\bbM} \longrightarrow \tilde{\bbM}_L \la{fibC}
 \end{equation}
 and one defines non-compact left chiral superspace in
 terms of this
 fibration as $\pi(\bbM)$.
 Right chiral superspace is defined similarly and has a single cross
 through the other odd node of the Dynkin diagram.

The spaces we will be concentrating on, however, will be harmonic
superspaces. Complexified $(N,p,q)$ harmonic superspace
$\tilde{\bbM}_H$
 has
 the following isotropy group
 \be
 \begin{picture}(300,200)(20,-100)
 \put(10,0){ $ \left( \ba{cc|ccccccccc|cc}
\bt &\bt & & &&&&&&&&&
\\ \bt &\bt & & &&&&&&&&&\\ \hline \bt &\bt &\bt &.
 &\bt&&&&&&&&\\ . &. &. &. &. &&&&&&&&\\ \bt &\bt &\bt &.
 &\bt&&&&&&&&\\ \bt &\bt &\bt &. &\bt &\bt &. &\bt &&&&&\\
 .&.&.&.&.&.&.&.&&&&&\\ \bt &\bt &\bt &. &\bt &\bt &. &\bt &&&&&\\ \bt
 &\bt &\bt &. &\bt &\bt &. &\bt&\bt&.&\bt&&\\
 .&.&.&.&.&.&.&.&.&.&.&&
\\ \bt &\bt &\bt &. &\bt &\bt &.
 &\bt&\bt&.&\bt&&\\ \hline
\bt &\bt &\bt &. &\bt &\bt &.
 &\bt&\bt&.&\bt&\bt&\bt\\
\bt &\bt &\bt &. &\bt &\bt &.
 &\bt&\bt&.&\bt&\bt&\bt\\ \ea \right) $ }
\put(80,35) {$
 \left. \phantom{\begin{array}{c} \bt\\ .
 \\ \times \end{array} } \right\}{\scriptscriptstyle p} $}
 \put(207,-40) {$ \left. \phantom{\begin{array}{c} \times \\ . \\ \bt
 \\ \end{array} } \right\}{\scriptscriptstyle q} $} \put(120,0) {$
 \left. \phantom{\begin{array}{c} \times \\ . \\ \bt
 \end{array} } \right\}{\scriptscriptstyle N-p-q} $}
 \end{picture}\la{harmonic}
 \ee
 The corresponding
Dynkin diagram is
\be
\begin{picture}(300,0)(-5,0)
\put(0,0){\makebox[0pt][l]{$\bt\hspace{1.5em}\otimes\hspace{1.5em}\bt\hspace{1.5em}
\times\hspace{1.5em}\bt
\hspace{1.5em}\times\hspace{1.5em}\bt\hspace{1.5em}\otimes
\hspace{1.5em} \bt$} \rule[.5ex]{4.5em}{.1ex} $\hspace{.4em}
\cdots \hspace{1em} \cdots \hspace{1em} \cdots \hspace{1em}\cdots
\hspace{.4em}$ \rule[.5ex]{4.5em}{.1ex}
 }
\end {picture}
\ee where the middle crosses are on the $p$th and $(N-q)$th
central nodes.
 Again one has a fibration this time with Minkowski space as the base
 manifold
 \begin{equation}
 \tilde{\bbM}_H \longrightarrow \tilde{\bbM} \la{fibH}
 \end{equation}
 and so in the non-compact version we define $\bbM_H=\pi^{-1}(\bbM)$
 which has the form of complex super
 Minkowski space times an internal flag space.

The related
 $(N,p,q)$ analytic superspace has the same body but fewer odd
 coordinates. It has the following isotropy group
 \be
 \begin{picture}(300,200)(20,-100)
 \put(10,0){ $ \left( \ba{cc|ccccccccc|cc}
\bt &\bt &\bt &. &\bt&&&&&&&&
\\ \bt &\bt &\bt &. &\bt&&&&&&&&\\ \hline \bt &\bt &\bt &.
 &\bt&&&&&&&&\\ . &. &. &. &. &&&&&&&&\\ \bt &\bt &\bt &.
 &\bt&&&&&&&&\\ \bt &\bt &\bt &. &\bt &\bt &. &\bt &&&&&\\
 .&.&.&.&.&.&.&.&&&&&\\ \bt &\bt &\bt &. &\bt &\bt &. &\bt &&&&&\\ \bt
 &\bt &\bt &. &\bt &\bt &. &\bt&\bt&.&\bt&\bt&\bt\\
 .&.&.&.&.&.&.&.&.&.&.&.&.
\\ \bt &\bt &\bt &. &\bt &\bt &.
 &\bt&\bt&.&\bt&\bt&\bt\\ \hline
\bt &\bt &\bt &. &\bt &\bt &.
 &\bt&\bt&.&\bt&\bt&\bt\\
\bt &\bt &\bt &. &\bt &\bt &.
 &\bt&\bt&.&\bt&\bt&\bt\\ \ea \right) $ }
\put(80,35) {$
 \left. \phantom{\begin{array}{c} \bt\\ .
 \\ \times \end{array} } \right\}{\scriptscriptstyle p} $}
 \put(207,-40) {$ \left. \phantom{\begin{array}{c} \times \\ . \\ \bt
 \\ \end{array} } \right\}{\scriptscriptstyle q} $} \put(120,0) {$
 \left. \phantom{\begin{array}{c} \times \\ . \\ \bt
 \end{array} } \right\}{\scriptscriptstyle N-p-q} $}
 \end{picture}\la{analytic}
 \ee
with corresponding Dynkin diagram
\be
\begin{picture}(300,0)(-10,0)
\put(0,0){\makebox[0pt][l]{$\bt\hspace{1.5em}\ominus\hspace{1.5em}\bt\hspace{1.5em}
\times\hspace{1.5em}\bt
\hspace{1.5em}\times\hspace{1.5em}\bt\hspace{1.5em}\ominus
\hspace{1.5em} \bt$} \rule[.5ex]{4.7em}{.1ex} $\hspace{.2em}
\cdots \hspace{1em} \cdots \hspace{1em} \cdots \hspace{1em}\cdots
\hspace{.4em}$ \rule[.5ex]{4.5em}{.1ex}
 }
\end {picture}
\ee
 This space has only $(N-p)$ $\th$'s and $(N-q)$ $\vf$'s.
 There is a double fibration involving compactified super Minkowski
 space, harmonic
 superspace and analytic superspace
 \begin{equation}
 \begin{picture}(300,120)(-80,-10)
 \put(70,80){$\tilde{\bbM}_H$} \put(60,68){\vector(-1,-1){50}}
 \put(80,68){\vector(1,-1){50}} \put(-10,0){$\tilde{\bbM}_A$}
 \put(60,0){$\Longleftrightarrow$} \put(130,0){$\tilde{\bbM}$}
 \put(11,50){$\pi'$} \put(111,50){$\pi$}
 \end{picture}\la{fibA}
 \end{equation}
and non-compact analytic superspace is defined as $\bbM_A=\pi'(\bbM_H)$.
Generalised $(N,p,q)$ spaces can be defined, which have the same
number of $\th$'s and $\vf$'s as $(N,p,q)$ space, but have a
different internal space. These are given by the same Dynkin
diagram as above, but with any number of extra crosses inserted
between the two already there.

We note here also that the supertwistor space introduced by Ferber~\cite{Ferber:1978qx}
 corresponds to the Dynkin diagram
\be
\begin{picture}(300,20)(-10,-10)
\put(0,0){\makebox[0pt][l]{$\times\hspace{1.5em}\ominus\hspace{1.5em}
    \bt \hspace{1.5em}
\bt\hspace{1.5em}\cdots
\hspace{1.5em}\bt\hspace{1.5em}\bt\hspace{1.5em}\ominus\hspace{1.5em}
\bt$} \rule[.5ex]{6.9em}{.1ex} $\hspace{4.5em}$
\rule[.5ex]{6.9em}{.1ex}
 }
\end {picture} \label{stwistor}
\ee

\subsection{Superconformal representations}

Weights, raising and lowering operators and highest weights are
defined as in the bosonic case. Weights $\l_i$ dual to the simple
roots are defined as in~(\ref{li}) so that any weight can be
expressed uniquely as
 \be
  \l=\sum_i (\l,\a_i^{\vee})\l_i.\label{lambda}
  \ee
  A weight is then represented on a Dynkin diagram by putting the
 label $n_j=(\l,\a_j^{\vee})$ over the $j$-th node.

For the Lie superalgebras $\gs \gl(r|s)$ (for $r>s \geq 1$) the
representations with highest weight $\l$ are finite dimensional if
and only if the even \textit{distinguished} Dynkin coefficients
$n_{i}^D$ are positive integers~\cite{Cornwell}. This is the
equivalent of the dominant integral condition in the bosonic case,
the difference now being that the odd Dynkin coefficient can be
any (complex) number. This is important for representing
operators with non-integral dilation weight as superfields on analytic
superspaces.

Irreducible highest weight representations of $SL(4|N)$ can
therefore be defined by giving $3+N$ Dynkin coefficients. Unitary
irreducible highest weight representations of the superconformal
group are usually defined by giving the following $3+N$ quantum
numbers: Lorentz spin, $J_1, J_2$, dilation weight, $L$, R-charge
$R$, and the Dynkin labels of the internal group, $a_1 \ldots
a_{N-1}$. These can be related to the Dynkin coefficients. In
order to work out this relation, we first calculate
$H_i\in\gh$ dual to the simple co-roots $\a_i^{\vee} \in \gh^*$. For the choice
of simple roots $S$~(\ref{S}) using the new basis~(\ref{eq:g}) we
find that \be \ba{rcl}
H_1 &=&\mbox{diag}(1,-1|0,\dots0|0,0)\\
H_2 &=&\mbox{diag}(0,-1|-1,0,\dots,0|0,0)\\
H_3   &=&\mbox{diag}(0,0|1,-1,0,\dots,0|0,0)\\
\vdots&&\\
H_{N+1}&=&\mbox{diag}(0,0|0,\dots,1,-1|0,0)\\
H_{N+2}&=&\mbox{diag}(0,0|0,\dots,1|1,0)\\
H_{N+3}&=&\mbox{diag}(0,0|0,\dots,0|1,-1).\\
\ea \label{hvee}\ee For a representation with highest weight $\l$,
the Dynkin coefficients corresponding to the simple root $\a_i$
are given by $n_i = (\l, \a_i^{\vee}) = \l(H_i)$. The quantum
numbers $Q$ given above satisfy a similar equation $Q=\l(\hat{Q})$
where $\hat{Q} \in \gh$ are as follows \be \ba{ll}
\hat{J}_1={1\over2}\left(\ba{c|ccc|c}
 \s_3&& && \\
 \hline
  &&& & \\
  &&\phantom{{4\over N} 1_N}&&\\
  &&&&\\
\hline
 &&&& \ea
\right) &
 \hat{J}_2={1\over2}\left(\ba{c|ccc|c}
 && && \\
 \hline
  &&& & \\
  &&\phantom{{4\over N} 1_N}&&\\
  &&&&\\
\hline
 &&&&\s_3 \ea
\right)\\ \\
 \hat{L}={1\over2}\left(\ba{c|ccc|c}
 -1_2&& && \\
 \hline
  &&& & \\
  &&\phantom{{4\over N} 1_N}&&\\
  &&&&\\
\hline
 &&&&1_2 \ea
\right)&
 \hat{R}={1\over2} \left(\ba{c|ccc|c}
 1_2&& & \\
\hline &&&&\\ &&{4\over N}1_N&&\\ &&&&\\ \hline
 &&&& 1_2
 \ea \right)\ea \label{gen}
 \ee
and \be \hat{a}_i =H_{i+2} \qquad i=1 \dots N-1 . \ee We
thus find that
 \be \ba{lll} n_1 = 2J_1; \qquad &n_2 =
 {1\over2}(L-R) + J_1 + {m\over
 N} -m_1; \qquad &n_{2+i}=a_i \ \ (i=1 \dots N-1); \\
 n_{N+3}=2J_2; \qquad &n_{N+2} = {1\over2}(L+R)+J_2-{m\over N}; \ea \label{ni}
 \ee
 where \be m:=\sum_{k=1}^{N-1}k a_k;\qquad
m_1:=\sum_{k=1}^{N-1}a_k. \label{m}\ee
 Note that $m$ is the total number of
boxes in the $\gs\gl(N)$ Young tableau, while $m_1$ is the number
of boxes in the first row.

Finite dimensional representations of parabolic subgroups can be
represented on a Dynkin diagram with crosses through it. As in the
bosonic case an irreducible representation of $\gp=\gl \oplus \gn$
corresponds to an irreducible representation of $\gl$ as $\gn$
acts trivially. $\gl$ splits into a semisimple part $\gl_S$ and
its centre $\gl_Z$ and thus a representation of $\gp$ can be
specified by giving the highest weight of the representation of
$\gl_S$ together with some numbers giving $\com$-weights (the
representation of $\gl_Z$). Roughly speaking, on the Dynkin
diagram for $\gp$ the Dynkin coefficients above crossed nodes
correspond to the $\com$-weights and the remaining coefficients
give the representation of $\gl_S$.

\subsection{Induced representations of supergroups}

Proceeding exactly as in the bosonic case we now describe
representations of a complex simple Lie supergroup G as fields on
supercoset spaces $P\bsh G$ which carry a representation of the
parabolic subgroup $P$. We represent these by using the Dynkin
diagram for the representation of $P$ which the fields carry. For
example the Wess-Zumino multiplet has quantum numbers $L=1$,
$R=-1$, $J_1 = J_2=0$ and thus, from \eq{ni}, has Dynkin diagram
\be
\begin{picture}(30,10)(0,0) \put(0,0){\makebox[0pt][l]{$\bt
\hspace {2em}
 \ominus \hspace{2em}\ominus \hspace{2em}\bt$} \rule[.5ex]{8.3em}{.1ex} }
 \put(0,10){\tiny 0 \hspace{3em} 1 \hspace {3.8em}0 \hspace{2.8em} 0}
 \end{picture}
\ee It can be represented as a chiral field (i.e. a field $A$
satisfying the constraint $\bar{D}_{\a}A=0$) on (real) $N=1$ super
Minkowski space. In complex superspace this field is most
naturally given as an unconstrained scalar field on $N=1$ chiral
superspace~(\ref{chiral}). Its Dynkin diagram is \be
\begin{picture}(30,10)(0,0) \put(0,0){\makebox[0pt][l]{$\bt
\hspace {2em}
 \otimes \hspace{2em}\ominus \hspace{2em}\bt$} \rule[.5ex]{8.3em}{.1ex} }
 \put(0,10){\tiny 0 \hspace{3em} 1 \hspace {3.8em}0 \hspace{2.8em} 0}
 \end{picture}
\ee
The fact that there are no non-zero coefficients above uncrossed
nodes tells us that this is a scalar superfield.

The $N=2$ hypermultiplet has quantum numbers $L=1$, $R=0$,
$J_1=J_2=0$, $a_1=1$ and thus has Dynkin diagram
\be
\begin{picture}(30,10)(0,0) \put(0,0){\makebox[0pt][l]{$\bt
\hspace {2em}
 \ominus \hspace{2em}\bt \hspace{2em}\ominus \hspace{2em}\bt$} \rule[.5ex]{11.1em}{.1ex} }
 \put(0,10){\tiny 0\hspace{3.5em} 0 \hspace{3.2em} 1 \hspace {3.3em}0
 \hspace{3em} 0}
 \end{picture}
\ee
It can
be given as a field on (real) super Minkowski space, $W_i$, carrying an
$SU(2)$ index, and satisfying the constraint \eq{DW}
\be
\ba{rcl}
\bar{D}^i_{\adt}W_j &=& \frac12 \d^i_j \bar{D}^k_{\adt}W_k\\
D_{\a i}W_j &=& D_{\a [i}W_{j]}. \ea
\ee
This has Dynkin diagram
\be
\begin{picture}(30,10)(0,0) \put(0,0){\makebox[0pt][l]{$\bt
\hspace {2em}
 \otimes \hspace{2em}\bt \hspace{2em}\otimes \hspace{2em}\bt$} \rule[.5ex]{11.5em}{.1ex} }
 \put(0,10){\tiny 0\hspace{3.5em} 0 \hspace{3.2em} 1 \hspace {3.3em}0
 \hspace{3em} 0}
 \end{picture}
\ee
Note that this has a 1 above the central node, confirming that the
field has a single $SU(2)$ index.
Another
alternative description of this multiplet is as a field on
analytic space. Here it is an unconstrained field $W$ without indices,
and has Dynkin diagram
\be
\begin{picture}(30,10)(0,0) \put(0,0){\makebox[0pt][l]{$\bt
\hspace {2em}
 \ominus \hspace{2em}\times \hspace{2em}\ominus \hspace{2em}\bt$} \rule[.5ex]{11.5em}{.1ex} }
 \put(0,10){\tiny 0\hspace{3.5em} 0 \hspace{3.5em} 1 \hspace {3.3em}0
 \hspace{3em} 0}
 \end{picture} \label{hyper}
\ee

The above examples are well known and all transform trivially under
supergroups. To include all UIRs on a particular space it is necessary
to consider fields which transform non-trivially under supergroups.

\subsection{Unitary bounds}\label{ubounds}

Unitary irreducible highest weight representations of the superconformal group
satisfy the following unitary bounds
\cite{Flato:1984te,Dobrev:1985qv,Dobrev:1987qz,Binegar:1986ib,Morel:1985if}:
\be
\ba{rrclrcl}
A)& L&\geq&2+2J_2-R+{2m \over N}, &L&\geq& 2 + 2J_1+R+2m_1-{2m\over N}\\
B)& L&=&-R +{2m \over N}, &L &\geq&1+m_1 +J_1, \qquad J_2=0\\
C)& L&=&m_1, &R&=&{2m \over N}-m_1, \qquad J_1=J_2=0
\ea\label{abc} \ee
 (or $J_1\rightarrow
  J_2$, $r\rightarrow -r$, ${2m\over
N}\rightarrow 2m_1-{2m\over N}$ for series B. In terms of the
Dynkin labels these bounds have the following form: \be
\ba{rrclrcl}
A)& n_2&\geq &n_1 +1,& n_{N+2}&\geq&n_{N+3}+1\\
B)& n_2&\geq &n_1 +1,& n_{N+2}&=&0, \qquad n_{N+3}=0\\
C)& n_2&=&0,&n_{N+2}&=&0, \qquad n_1=n_{N+3}=0 \ea \label{bounds}
\ee (or $n_1 \rightarrow n_{N+3}$, $n_2 \rightarrow n_{N+2}$ for
series B.)

 Notice that for series C the coefficients above the two odd
 nodes, and above the first and last nodes must be zero, and for series
 B one either has the first two coefficients zero, or the last two
 zero. We will find that for
 fields on analytic superspaces there is a simple interpretation for
 these bounds, namely that for fields which carry
 non-trivial representations of supergroups, all superindices are downstairs.

 \section{Reducibility conditions and constraints}
 \la{sect:red}

Having indicated the way in which representations of the
superconformal group can be carried by superfields on different
spaces, we come to the question of when these representations are
irreducible. We know that on super Minkowski space the superfields
often have to satisfy constraints in order to be irreducible. In
this section we review the constraints that UIRs must satisfy on
Minkowski and super Minkowski space, and show why no such
constraints are required on $(N,p,q)$ analytic superspaces with
$p,q \geq 1$. In other words superfields on such analytic
superspaces are automatically irreducible.

 \subsection{Verma modules and invariant submodules}\la{verma}

 The question of when representations are reducible can be answered
 with the help of Verma modules (see~\cite{Baston}). A Verma module $V(\l)$ with
 highest weight $\l$ is the set spanned by the highest
 weight vector
 and all its descendants. A positive root $\a$ has a corresponding
 co-root, $\a^{\vee}$
 (\ref{co-root}), and corresponding Cartan
 matrix $H_{\a}=h_{\a^\vee}$ (\ref{ha}). It also has an associated
 raising (lowering) operator, $e_{\pm \a}$ defined by
 \be
 [h,e_{\pm \a}]=\pm \a(h)e_{\pm \a}
 \ee
 for all elements of the Cartan algebra $h$.
 A basis for the Verma module $V(\l)$ is given by applying all possible
 lowering operators on the highest weight state
 \be
 \prod_{\a \in \D^+}(e_{-\a})^{m_{\a}}|\l \rangle \la{mu}
 \ee
 where $\D^{+}$ is the set of positive roots and $m_{\a}$ are positive
 integers.
 This gives a vector of weight
 \be
 \l - \sum m_{\a} \a.
 \ee
 This module is reducible if and only if it contains a vector,
 called a singular vector, which has the
 characteristic of the highest weight state of another Verma
 module. In other words we are looking for vectors
 $|\m\rangle$ in $V(\l)$ which are annihilated by all raising operators
 $e^+$ and are thus highest weight state of invariant submodules.

 We give some motivations for the formula for finding invariant submodules
 for $\gs \gl(n)$. We have
 $n-1$ positive simple roots,
 $\a_j$,
 with corresponding Cartan matrices $H_j$ and corresponding raising
 and lowering operators $e^+_j$ and $e^-_j$. These are
 given by
 \bea
 H_j&=& \hat{e}_{jj}-\hat{e}_{(j+1)(j+1)}\\
 e^+_j&=& \hat{e}_{j(j+1)}\\
 e^-_j&=& \hat{e}_{(j+1)j}
\eea
 where $\hat{e}_{jk}$ is the matrix with zeros everywhere except in
 the $(jk)$ entry where there is a 1.
 First consider vectors of the form
 \be
 |\m\rangle = (e^-_j)^n|\l\rangle
 \ee
 with weight
 \be
 \m=\l-n \a_j.
 \ee
 This
 is annihilated by all the operators $\{e_i^+: i\neq j \}$, since
 $e_i^+$ commutes with $e_j^-$ for $i \neq j$. One can show that
 \be e_j^+ |\m\rangle = n(n_j-n+1)|\m\rangle
 \ee
 where $n_j=(\l, \a^{\vee}_j)$ is the $j$th Dynkin label for the weight
 $\l$.
 If we thus choose $n=n_j+1$ then $|\m\rangle$ is annihilated
 by all raising operators and so $V(\m)$
 is an invariant submodule of $V(\l)$.

 To find all possible submodules of $V(\l)$ we need the concept of a
 Weyl reflection.
 For each
 root $\a$ a
 Weyl reflection $\s_{\a}$ acts on the weight space as
 \be
 \s_{\a}(\l)=\l-(\l,\a^{\vee})\a.\la{Weyl}
 \ee
 Then the highest weight of the submodule defined above,
 $\m=\l - (n_j + 1)\a_j $, is just
 \be
 \m = \s_{\a_j}(\l+\r)-\r \qquad (\l+\r,\a^{\vee})>0
 \ee
 where $\r$ is the weight with Dynkin coefficients $(1,\dots,1)$. It is
 clear that if we repeat this process we will obtain the highest
 weight of another invariant submodule
 $\m'=\s_{\a_k}\s_{\a_j}(\l+\r)-\r$, where $\a_i$, $\a_j$ are simple roots.

 In fact the highest weights of all invariant submodules $V(\m)$ can
 be obtained
 by repeated Weyl reflections (see~\cite{Baston}). That is, all highest weights of
 invariant submodules have the form
 \be
 \m = \s_{\a_M}\s_{\a_{M-1}}\dots \s_{\a_1}(\l+\r)-\r
 \ee
 where ${\a_p}$ is a sequence of simple positive roots satisfying
 \be
 m_p \equiv (\s_{\a_{p-1}}\s_{\a_{p-2}}\dots \s_{\a_1}(\l
 +\r),\a_p^{\vee}) > 0 \qquad p=1 \dots M
 \ee
 The weight vector is given in terms of the $m_p$ as
 \be
 \m= \l - \sum m_p \b_p.
 \ee

\subsection{Reducibility conditions on Minkowski space}

 Induced representations on non-compact manifolds may also posses invariant
 subrepresentations. These are essentially the same as above with the caveat
 that many of the invariant subrepresentations found previously will be automatically zero
 because of the construction of induced representations. To find the remaining subrepresentations
 we proceed as follows (the differential operators which one must apply to make
 irreducible representations of the real conformal group $SU(2,2)$ were found in
 \cite{Dobrev:1977qv,Dobrev:1978fv,Dobrev:1985ts}). On Minkowski
 space we say that a positive root $\b$ is non-compact if
 $\gg_{\b}\subset \gn$ where
 \be
 \gn=
 \left\{ \left(\ba{cccc}   &
  &&\\   &
  &&\\
  \bt & \bt
 &  &  \\
  \bt & \bt
 &  &
 \ea
 \right)\right\}
 \ee
 Then the criteria for finding invariant submodules given in the
 previous subsection is adapted to the following.

 A representation with highest weight $\l$ is reducible if for
 some non-compact root $\b$ one of the following two conditions holds:
 \bea
 1.) &\qquad& (\l+\r,\b^{\vee})>0 \la{cond} \\  2.) &\qquad & \l'=\s_{\b}(\l + \r) - \r
 \ \ \mbox{ is dominant integral for }\gp.
 \eea
  If the first condition is met but
 not the second, then the representation may still be
 reducible if $\l'=\s_{\b}(\l + \r) - \r$ satisfies these two
 conditions. If $\l'$ satisfies the first condition for a
 non-compact root $\b'$ but not the second condition, then we
 repeat this process, defining $\l''=\s_{\b'}(\l' + \r) - \r$ and
 seeing whether it satisfies any of the conditions.

 Note that the first
 condition tells us that a state with weight
 $\l'=\s_{\b}(\l + \r) - \r$ can, in principle, be obtained by
 applying lowering
 operators ($\l' \prec \l$ using the partial ordering defined
 in the appendix (\ref{po})).
 In the cases we consider this state is annihilated by
 all raising operators and thus it is the highest weight state of an
 invariant subrepresentation, $V(\l')$ of $V(\l)$ (see previous subsection).
 When
 considering representations as fields on a parabolic space some of
 these subrepresentations are zero automatically, namely those for which $\l'$ is not
 dominant with respect to $\gp$. This is because we demand that the fields
 carry irreducible finite dimensional representations of the parabolic
 subgroup. Hence we require the second
 condition.
 Note also that we only need to consider applying non-compact
 roots because if we apply compact roots then again the
 subrepresentation is zero automatically since we are considering
 irreducible representations of the parabolic subgroup.

 We make our representations irreducible by demanding that all
 invariant subrepresentations are zero.
 The irreducibility conditions are therefore given by
 differential operators which take $V(\l) \mapsto V_{\l'}$ (or $V_{\l}
 \mapsto V_{\l''}$ etc.). Setting these maps to zero will give
 irreducible representations.

 For Minkowski space there are four non-compact roots $\a_2, \a_2 +\a_1,
 \a_2+\a_3, \a_1 +\a_2 +\a_3$ and condition 1) says that in order for
 a representation to be reducible one of the following 4 conditions
 must hold
 \bea
 n_2 +1&>& 0 \label{1}\\
 n_1 + n_2 +2 &>&0 \label{2}\\
 n_2 + n_3 +2 &>&0 \label{3}\\
 n_1 +n_2+n_3+3&>&0 \label{4}.
 \eea
 If we put these conditions together with the unitary bounds
 (see appendix (\ref{bbounds}) we
 find that the only reducible representations are those with the
 following Dynkin labels
 \be \ba{c} (0,0,0)\\ (k,-k-1,0)\\(0,-k-1,k)\\(n_1,-n_1-n_3-2,n_3) \ea
 \ee
 where $k,n_1,n_3$ are non-negative integers.

 Consider the
 representation with highest weight $\l=(k,-k-1,0)$. We find that
 $\l'=\s_{\a_1 +\a_2}(\l+\r)-\r=(k-1,-k-2,1)$ which is dominant
 integral for $k\geq0$. Thus the reducibility condition is given by
 the map
\be
  \begin{picture}(60,10)(0,0) \put(0,0){\makebox[0pt][l]{$\bt
 \hspace {2em}
 \xz \hspace{2em}\bt$} \rule[.5ex]{5.5em}{.1ex} }
 \put(0,10){\tiny k \hspace{2.2em} -k-1 \hspace {2.5em}0}
 \end{picture}
 \qquad
 \longrightarrow \qquad
 \begin{picture}(60,10)(0,0) \put(0,0){\makebox[0pt][l]{$\bt
 \hspace {2em}
 \xz \hspace{2em}\bt$} \rule[.5ex]{5.5em}{.1ex} }
 \put(0,10){\tiny k-1 \hspace{1.5em} -k-2 \hspace {2.5em}1}
 \end{picture}
 \ee
 In terms of fields we have $\psi_{\a_1 \dots \a_k}$ with the
 irreducibility condition
 \be
 \del^{\a_1}_{\bdt}\psi_{\a_1 \dots \a_k}=0
 \ee
  and we recognise the usual
 massless field equations for right handed fields. In a similar way
 the representation with highest weight $(0,-k-1,k)$ gives left-handed
 massless fields.

 The case $k=0$ is more complicated as $\l'=(-1,-2,0)$ is not dominant
 for $\gp$ so condition 2) is not satisfied. We therefore have to
 apply the same procedure to $\l'$. We find that $\l'$ satisfies
 conditions 1) and 2) for the non-compact root $\a_2 +\a_3$ giving
 $\l''=(0,-3,0)$. The reducibility condition is therefore given by
 \be
 \begin{picture}(60,10)(0,0) \put(0,0){\makebox[0pt][l]{$\bt
 \hspace {2em}
 \xz \hspace{2em}\bt$} \rule[.5ex]{5.5em}{.1ex} }
 \put(0,10){\tiny 0 \hspace{2.5em} -1 \hspace {3em}0}
 \end{picture}
 \qquad
 \longrightarrow \qquad
 \begin{picture}(60,10)(0,0) \put(0,0){\makebox[0pt][l]{$\bt
 \hspace {2em}
 \xz \hspace{2em}\bt$} \rule[.5ex]{5.5em}{.1ex} }
 \put(0,10){\tiny 0 \hspace{2.5em} -3 \hspace {3em}0}
 \end{picture}
 \ee
 As a field we have a scalar field $\phi$ satisfying the second order
 equation
 \be
 \del_{\a \adt} \del^{\a \adt} \phi =0
 \ee
  and we recognise the usual massless scalar equation.

 For the representations $(n_1,-n_1-n_3-2,n_3)$ we find fields
 $\psi_{\a\dots\c \adt \dots \cdt}$ with $n_1$ undotted and $n_3$
 dotted indices, satisfying the equation
 \be
 \del^{\a \adt} \psi_{\a \b\dots\c \adt \bdt \dots \cdt} =0
 \ee
 if $n_1,n_3 > 0$. In the case $n_1=0$ or $n_3=0$ we find $\l'$
 does not satisfy condition 1) and thus the representations are
 irreducible (without any conditions).

 \subsection{The supersymmetric case}

 We briefly adapt some of the discussion in section \ref{verma} to
 the supergroup $\gs \gl(4|N)$ following~\cite{Dobrev:1987qz}. Again, for each positive root $\a$ we have
 a corresponding raising (lowering) operator $e_{\pm \a}$ defined by
 \be
 [h,e_{\pm \a}]=\pm \a(h)e_{\pm \a}
 \ee
 for all elements of the Cartan algebra $h$. The Verma module
 $V(\l)$ is obtained by applying all possible combinations of
 lowering operators to the highest weight state and this module is reducible
 if and only if it contains an invariant
 submodule
 $V(\m)$ with highest weight $\m$. These are known as singular
 vectors. So we are looking for vectors
 $|\m\rangle$ in $V(\l)$ which are annihilated by all raising operators
 $e^+$.

 Again we will motivate the discussion by considering simple
 roots. For even simple roots we obtain, as
 previously, that the weight $\m=\l - (n_j + 1)\a_j $ is the highest weight
 of an invariant submodule.
 For odd simple roots $\b$, consider the state
 $|\m\rangle=e^-_{\b}|\l\rangle$ (note we can only apply the lowering
 operator once as it is odd.) Then
 \be
 e^+_{\b}|\m\rangle=H_{\b}|\l\rangle.
 \ee
 So this is zero if and only if $H_{\b}|\l\rangle=0$.

 We define Weyl reflections for superalgebras. For even roots
 they are defined as in \eq{Weyl} whereas
 for odd roots $\b$ we define
 \be
 \s_{\b}(\l)=\l - \b , \qquad (\b,\b)=0.
 \ee
 We also define the weight $\r$ to have Dynkin
 coefficients $(1,0,1,1,\dots,1,0,1)$ so it has 1's above all even
 nodes and 0's above the odd nodes in the Dynkin diagram.
 We see that the invariant submodules have highest weights $\m$
 where
 \be
 \m = \s_{\a}(\l+\r)-\r \qquad (\l + \r,
 \a^{\vee})>0 \quad (\a \mbox{ even}) \qquad (\l + \r,
 \a^{\vee})=0 \quad (\a \mbox{ odd}).
 \ee
 Once again, all invariant submodules are obtained by repeated
 Weyl reflections.

 \subsection{Reducibility constraints for fields on super
 Minkowski space}\la{sired}

 In this section we review the constraints fields on super
 Minkowski space or harmonic superspace must satisfy in order to carry
 irreducible
 representations of the superconformal group.

 We define non-compact positive roots $\b$ (even or odd) to be
 those such that
 $\gg_{\b} \subset \gn$ where $\gn$ is the set of matrices of the form
 \be
 \left( \ba{cc|ccc|cc}
                           &   &&&&&\\
                           &    &&&&&\\
\hline                   \bt  &  \bt  &   & &  &&\\
                        . &.  & & & &&\\
                         \bt  &  \bt  &  & &   &&\\
\hline                   \bt  & \bt  & \bt  &.  & \bt  &   &  \\
                         \bt  & \bt  & \bt  &.  & \bt &  &
                        \ea \right)
\ee
 The reducibility conditions in the bosonic case \eq{cond} are
 straightforwardly
 modified for the supersymmetric case \cite{Dobrev:1987qz}. Field representations on super
 Minkowski space with highest weight $\l$ are reducible if for some $\b$, the
 following two conditions are met:
 \bea
 1.) &\qquad&(\l+\r,\b^{\vee})=0 \ (\b \mbox{ odd }) \qquad
 (\l+\r,\b^{\vee})>0 \ (\b \mbox{ even })  \\
 2.) &\qquad & \s_{\b}(\l + \r) - \r
 \ \ \mbox{ is dominant integral for }\gp.
 \eea
 The
 weight $\l'=\s_{\b}(\l + \r) - \r$ is the highest weight of an
 invariant subrepresentation. If
 condition 1) is met but condition 2) is not then one checks to see if
 $\l'=\s_{\b}(\l + \r) - \r$ satisfies
 similar conditions. Repeat this process if $\l'$ only satisfies condition
 1). The conditions one must impose to make the representations
 irreducible are given by differential operators which map
 $V_{\l} \rightarrow V_{\l'}$ (or $V_{\l} \rightarrow V_{\l''}$ etc.)
 where $V_{\l}$ denote the representation with highest weight $\l$. The
 differential operators correspond to the infinitesimal Lie algebra
 element $\gg_{-\b}$ since $\gg_{-\b}$ takes a state with weight $\l$
 to one with weight $\l'$.

 So for super Minkowski space we have $4N$ non-compact odd roots and 4
 non-compact even roots so there are $4N+4$ possible reducibility conditions.
 Condition 1) becomes (for the $4N$ odd roots)
 \bea
 n_2 + \sum_{i=1}^{a} n_{2+i} +a &=&0 \la{one}\\
 n_2 + \sum_{i=1}^{a} n_{2+i} +a -n_1 -1&=&0\la{two}\\
 n_{2+N} + \sum_{i=1}^{a} n_{2+i} +a &=&0\la{three}\\
 n_{2+N} + \sum_{i=1}^{a} n_{2+i} +a - n_{3+N}-1&=&0\la{four}
 \eea
 where $a=0 \dots N-1$.
 Putting these together with the unitary bounds \eq{bounds} we find
 that the equations can only be satisfied for $a=0$ giving
 \be \ba{cccc}
 \mbox{root} \b & \mbox{condition} & \l & \l'\\
 \hline
 \a_2 & n_1=n_2=0 & \mbox{\tiny$(0,0,n_3,\dots,n_{N+3})$}& \mbox{\tiny$(1,0,n_3+1, n_4, \dots,
 n_{N+3})$} \\
 \a_1 + \a_2 & n_2=n_1 + 1 & \mbox{\tiny$(n_1, n_1 +1, n_3, \dots, n_{N+3})$}&
 \mbox{\tiny$(n_1-1, n_1, n_3+1, n_4,\dots, n_{N+3})$}\\
 \a_{N+2}& n_{2+N}=n_{3+N}=0 &
 \mbox{\tiny$(n_1,\dots,n_{N+1},0,0)$}&\mbox{\tiny$(n_1,\dots,n_{N},n_{N+1}+1,0,1)$}
 \\
 \a_{N+2}+\a_{N+3} & n_{2+N}=n_{3+N}+1 &
 \mbox{\tiny$(n_1,\dots,n_{N+1},n_{N+3}+1,n_{N+3})$}&
 \mbox{\tiny$(n_1,\dots,n_{N+1}+1,n_{N+3},n_{N+3}-1)$}.
 \ea \la{con}\ee
 Furthermore for $N>0$ the conditions coming from the even
 compact
 roots are incompatible with the unitary bounds and therefore these
 conditions above are the only possible shortening
 conditions. Corresponding to each of these conditions there are
 the
 following irreducibility conditions
 \bea
 \a_2 &: \qquad &(D_{\a i} W_{j k \dots})_{(n_3+1,n_4,\dots)}=0\\
 \a_1 + \a_2 &: \qquad & (D^{\a}_{i}  W_{ j k\dots l \ \a \dots
 \c})_{
 (n_3+1,n_4,\dots)} =0
 \eea
 where the subscript numbers tell us which $SU(N)$ representation to
 take.
 There are similar conjugate conditions for the other two cases. If
 $n_1=0$ the second equation does not apply. This is because
 condition 2) above is not met i.e. $\l'$  is not dominant for $\gp$
 and thus we have to consider reducibility conditions for $\l'$. One
 finds that the conditions 1) and 2) are met for $\b=\a_1$ giving
 \be \ba{ccc} \l' &\longrightarrow &\l''\\
      (-1, 0, n_3+1, n_4,\dots, n_{N+3}) & \longrightarrow &
 (0,0,n_3+2, n_4,\dots, n_{N+3}).
 \ea \ee
 This leads to the second order reducibility condition
 ($V_{\l}\rightarrow V_{\l''}$) for fields on
 Minkowski space
 \be
 D_{\a (i} D^{\a}_j W_{k)l\dots m}=0.
 \ee
 One gets a similar (complex conjugate) equation if $n_{3+N}=0$
 in the fourth reducibility equation.
 Note that all these results are consistent with the equations
 satisfied by the massless multiplets in section~\ref{massless}, and
 they agree with the results found
 in~\cite{Ferrara:2000eb} for fields on harmonic superspace.

 \subsection{Fields on analytic superspace}
 The irreducibility conditions lift up straightforwardly
 to $(N,p,q)$
 harmonic superspace under the fibration \eq{fibH}, since the fibres
 are compact (the
 Bott-Borel-Weil theorem can be applied on the fibres.) By this we
 mean that the reducibility condition remains $V_{\l} \rightarrow
 V_{\l'}$. The differential operators which implement this on
 harmonic superspace can be found in~\cite{Ferrara:2000eb} using a
 generalised harmonic superspace with internal space $B\bsh SL(N)$.

 If we now consider pushing these fields down to $(N,p,q)$
 analytic superspace under the fibration~\eq{fibA} (with $p,q
 \geq 1$) we find that the reducibility conditions are
 satisfied automatically. To see this we consider the respective
 isotropy groups of super Minkowski space, and analytic superspace
 (\ref{smink},\ref{analytic})
 \be
 \left( \ba{cc|ccccc|cc}
                         \bt  &  \bt  &\odot&&&&&&\\
                         \bt  &  \bt  &\odot&&&&&&\\
\hline                   \bt  &  \bt  & \bt&\bt  &. & \bt &\bt &&\\
                        \bt &\bt  &\bt &\bt &.&\bt&\bt &&\\
                          . &.  &. &.&.&. &. &&\\
                          \bt &\bt  &\bt &\bt &.&\bt&\bt &&\\
                         \bt  &  \bt  & \bt&\bt  &. & \bt &\bt &\odot&\odot\\
\hline                   \bt  & \bt  & \bt  &\bt  &.  & \bt  &\bt  & \bt  & \bt \\
                         \bt  & \bt  & \bt & \bt  &.  & \bt  & \bt  &
 \bt & \bt
                        \ea \right)
\qquad \qquad
\left( \ba{cc|ccccc|cc}
                      \bt  &  \bt  &\bt&&&&&&\\
                         \bt  &  \bt  &\bt&&&&&&\\
\hline                   \bt  &  \bt  & \bt&  & &  & &&\\
                        \bt &\bt  &\bt &\bt &.&\bt& &&\\
                          . &.  &. &.&.&. & &&\\
                          \bt &\bt  &\bt &\bt &.&\bt& &&\\
                         \bt  &  \bt  & \bt&\bt  &. & \bt &\bt &\bt&\bt\\
\hline                   \bt  & \bt  & \bt  &\bt  &.  & \bt  &\bt  & \bt  & \bt \\
                         \bt  & \bt  & \bt & \bt  &.  & \bt  & \bt  &
 \bt & \bt
                        \ea \right)
 \end{equation}
 In this diagram we use $(N,1,1)$ analytic
 superspace for illustration, but the result is true for any
 $(N,p,q)$ space
 with $p,q \geq 1$. The circles correspond to the subspaces
 $\gg_{-\b}$ for $\b\in\{ \a_2,\ \a_1 + \a_2,\ \a_{N+2},\
 \a_{N+2}+\a_{N+3}\}$, the
 roots associated with the reducibility conditions
 (\ref{con}). On super Minkowski space the infinitesimal Lie algebra elements
 $\gg_{-\b}$ lie in $\gn^-$, the generator of translations in super Minkowski
 space. And so the constraint $V_{\l}
 \rightarrow V_{\l'}$ is a differential operator.
 On any $(N,p,q)$ analytic superspace with $p,q\geq 1$ however, the
 subspaces $\gg_{-\b}$ lie in the isotropy group which means the
 map $V_{\l}
 \rightarrow V_{\l'}$ is no longer a differential operator but a
 rotation of the isotropy group. But since we are only
 considering finite irreducible representations of the isotropy group,
 the constraint $V_{\l'}=0$ will be solved automatically as otherwise
 $V_{\l'}$ would restrict to a subrepresentation of the isotropy group.

 We conclude that unitary irreducible representations of the
 superconformal group are given by unconstrained fields on
 analytic superspace.

\section{Fields with superindices}\label{sindices}

 We have argued that any representation of the
 superconformal group can be given as a holomorphic field on any
 superflag space (except supertwistor spaces). We will here give
 some explicit examples of fields which transform under
 supergroups in analytic superspace. We will see that they do indeed
 give irreducible representations.

\subsection{Superindices}\la{sin}

 The fields on $(N,p,q)$ analytic superspace will carry irreducible
 representations of the
 supergroups $\gs \gl(2|p)$ and $\gs \gl(2|q)$. All finite dimensional
 irreducible representations of $\gs \gl(N)$
 can be described as symmetrised tensor products of the fundamental
 representation. Young Tableaux provide a useful way of describing these
 symmetries.
 It is true for $\gs \gl(r|s)$ that all finite-dimensional representations with
 integer Dynkin labels can be obtained as tensor products of the
 fundamental (and anti-fundamental) representation~\cite{Bars:1983ep}. We will
 deal with other finite dimensional representations later. Let $g^A{}_B$
  be a matrix in $\gs \gl(r|s)$ so that $A=(\a|a)$, where $\a$ runs
  from 1 to $r$ and $a$ runs from $r+1$ to $r+s$, is a superindex. We define
   the following generalised symmetrisation rules for the superindices
  \be
  (AB) = \left\{ \ba{lr} (\a \b)& A=\a, B=\b\\
                            (\a b)&A=\a, B=b\\
                            (a\b)&A=a, B=\b\\
                            {}[a b]&A=a, B=b
\ea \right. \qquad
 [AB]= \left\{ \ba{lr} [\a \b]& A=\a, B=\b\\
                            {}[\a b]&A=\a, B=b\\
                            {}[a\b]&A=a, B=\b\\
                            (a b)&A=a, B=b
\ea \right.\ee
 Indices in the range $A=1,2$ are called `even' and those in the
 range $A=3,4, \dots,2+s$ are called `odd' since they swap with a
 minus sign.
 So irreducible representations of $\gs \gl(r|s)$ are carried by
 tensors $W^{AB\dots C}$ with $m$ $\gs \gl(r|s)$ indices which satisfy
 various symmetry properties indicated by a super Young Tableau with
 $m$ boxes \cite{Dondi:1981qs}. The super Young Tableau gives the
 symmetries in the usual
 way (rows corresponding to symmetrised indices and columns
 corresponding to antisymmetrised indices) except that the definition
 of (anti)symmetry is modified to generalised (anti)symmetry described
 above. We can
 convert this Young Tableau into a Dynkin diagram
 by finding the highest weight state as in the bosonic
 case. We will do this explicitly for the group
 $\gs\gl(2|s)$
 which we will
 use later.

 The highest weight state $W^{1\dots1\ 2\dots2\dots}$ is found by
 putting numbers
 in the boxes
 of the Young tableau as follows
 \be
 \setlength{\unitlength}{.3mm}
\begin{picture}(500,160)(0,0)
\put(20,140){\framebox(180,20){}}
\put(23,146){1 1 1 1 1}
\put(100,146){$\cdots$}\put(151,146){1 1 1 1 1}
\put(20,120){\framebox(120,20){}}\put(23,126){2 2 2
2}\put(70,126){$\cdots$}\put(100,126){2 2 2 2}
\put(20,0){\framebox(20,120){}}\put(23,60){$\ba{c}3\\3\\3\\\vdots\\3\\3\\3\ea$}
 \put(40,30){\framebox(20,90){}}\put(43,75){$\ba{c}4\\4\\\vdots\\4\\4\ea$}
\put(60,60){\framebox(20,60){}}
 \put(80,90){\framebox(20,30){}}\put(75,105){\tiny $\ba{c}s+2\\s+2\ea$}
\end{picture}
 \ee

Unlike in the purely bosonic case, however, contravariant and
covariant representations are inequivalent as there is no invariant
$\vare$ tensor. Contravariant tensors
(which correspond to downstairs indices) have highest weight
states given by \be \setlength{\unitlength}{.3mm}
\begin{picture}(500,160)(0,0)
\put(100,140){\framebox(180,20){}}\put(103,146){2 2 2 2 2}
\put(180,146){$\cdots$}\put(233,146){2 2 2 2 2}
\put(100,120){\framebox(120,20){}}\put(103,126){1 1 1
1}\put(150,126){$\cdots$}\put(180,126){1 1 1 1}
\put(20,0){\framebox(20,160){}}\put(15,80){\tiny $\ba{c}s+2\\\\s+2\\\\s+2\\\\s+2\\\\\vdots\\\\s+2\\\\s+2\\\\s+2\\\\s+2\ea$}
 \put(40,30){\framebox(20,130){}}\put(35,95){\tiny $\ba{c}s+1\\\\s+1\\\\s+1\\\\\vdots\\\\s+1\\\\s+1\\\\s+1\ea$}
\put(60,60){\framebox(20,100){}}
 \put(80,90){\framebox(20,70){}}\put(83,125){$\ba{c}3\\\vdots\\3\ea$}
\end{picture}
 \ee
 where there are $m_1$ 1's, $m_2$ 2's etc. Then from the
definition of the Dynkin coefficients $n_i$, and from~(\ref{hvee})
we obtain \be \ba{rcl}
m_{s+2}-m_{s+1}   &=&n_{s+1}\\
m_{s+1}-m_{s}   &=&n_{s}\\
&\vdots &   \\
m_4-m_3   &=&n_3\\
m_3+m_2   &=&n_2\\
m_2-m_1   &=&n_1.
\ea \la{DY}\ee
Since the highest weight state determines the irreducible
 representation of $\gs \gl (2|s)$ uniquely, we can have different
 diagrams which correspond to the same representation. In particular,
 given a diagram with $m_1 \neq 0$ we can instead take the diagram
 with $m_i$ replaced by $m'_i$, where
 \be \ba{rlc}
 m'_1&=&0 \\
 m'_2&=&m_2-m_1\\
 m'_3&=&m_3 + m_1 \\
     & \vdots& \\
 m'_{s+2}&=& m_{s+2}+m_1
 \ea \ee
 and this will have the same Dynkin coefficients and thus correspond to the
 same representation. Thus for representations of $\gs \gl (2|s)$ we
 can assume without any loss of generality that
 $m_1=0$. However, this is not the case when we are considering
 representations of $\gg \gl (2|s)$.

Note that if we are
given the Dynkin weights $n_i\geq 0$ we will have a contravariant tensor
representation provided $m_3 \geq 1$ or $m_2=m_3=0$ (these are simply
conditions for the Young Tableau to have the correct shape). In terms
of Dynkin coefficients this is the condition
\be
n_2 \geq n_1 + 1  \mbox{ or } n_1=n_2=0.\label{condition}
\ee
On comparing this condition with the unitarity
 conditions~(\ref{bounds}) one can see that fields on analytic
 superspace will
 carry contravariant representations.

 There is some choice in the way the contravariant tensors
 transform. A field $W^A$ in the fundamental representation
 transforms as $W^A \mapsto g^A{}_B W^B$. Then a contravariant
 tensor $V_A$ can be chosen to transform so that either $V_A W^A$
 is invariant or $V^A W_A$ is invariant (in general they won't both be
 invariant). In the first case $V_A \mapsto V_B (g^{-1})^B{}_A$ as one might
 expect. In the second case however
 \be
 V_A \mapsto V_B(g^{-1})^B{}_A (-1)^{(A+B)}
 \ee
 where $(-1)^{(A+B)}$ is 1 if both A and B are even or if both are
 odd, and is -1 otherwise.

\subsection{Quasi-tensors}\la{quasi}

 Recall that finite dimensional representations of simple supergroups
 may have non-integral numbers above the odd nodes of the Dynkin
 diagram. It is not possible to accommodate
 such representations as ordinary tensors since these must clearly
 have integer Dynkin coefficients from~\eq{DY}. One must introduce the
 concept of `quasi-tensors'~\cite{Heslop:2001gp} to accommodate these
 representations.

 Consider representations of $\gs \gl(2|s)$ with Dynkin coefficients
 $[n_1, \dots, n_{s+1}]$. As mentioned, from unitarity we are only
 interested in representations with downstairs superindices, whose
 Dynkin labels
 satisfy $n_2 \geq n_1+1$ or $n_1 =n_2 = 0$~\eq{condition}. In fact, if we fix $n_1,n_3,n_4, \dots,
 n_{s+1}$ and vary $n_2$ we find that all representations for which $n_2 > n_1
 +1$ have the same dimension. These are called typical representations
 in the literature~\cite{Cornwell}. The representations with $n_2 =
 n_1 +1$ and $n_1 = n_2 = 0$ are known as `atypical' representations
 and have a smaller dimension (the latter case being the trivial one
 dimensional case). The easiest way to see this is to consider the
 representations acting on holomorphic sections of the parabolic space
 obtained by putting a cross through the single odd node of the Dynkin
 diagram for $\gs \gl(2|s)$:
\be
\begin{picture}(135,10)(0,0) \put(0,0){\makebox[0pt][l]{$\bt
\hspace {2em}
 \otimes \hspace{2em}\bt \hspace{2em}\cdots \hspace{2em}\bt$}
 \rule[.5ex]{7em}{.1ex} $\hspace{2.5em}$ \rule[.5ex]{1.5em}{.1ex} }
 \put(0,10){\tiny $n_1 \hspace{3.5em} n_2 \hspace{3.2em} n_3 \hspace {3.3em}
 \hspace{3em} n_{s+1}$}
 \end{picture}.
\ee
 This space has purely odd coordinates. The typical representations
 are given by unconstrained tensor superfields on this space (and
 hence are all of the same dimension for given Dynkin labels $n_1, n_3,
 \dots, n_{s+1}$), whereas
 the atypical representations must satisfy constraints in order to be
 irreducible. These are similar to the constraints found for
 representations on super Minkowski space in section~\ref{sired} and
 indeed the proof goes through using Verma modules in a similar way.

 Now given that these atypical representations all have the same
 dimension, we can represent them abstractly by the simplest possible
 such tensor, together with a real number to give the value of the odd
 Dynkin label. Specifically the $\gs \gl(2|s)$ representation with Dynkin labels
 $[n_1, n_1 +2 +p, n_3, \dots, n_{s+1}]$ we represent by the ordinary tensor
 specified by labels $[n_1, n_1+2, n_3, \dots, n_{s+1}]$ (we can
 find the corresponding tensor by following the rules in section~\ref{sin}) together with
 the number $p$. Since we know how this transforms for all
 integer values of $p \geq 0$, we can then continue $p$ to real (or
 even complex) values
 to find out how the quasi-tensors transform.

 The simplest example which illustrates this is given in $\gs
 \gl(2|1)$~\cite{Heslop:2001gp}. Consider representations with Dynkin labels $[0,
 p+2]$. For all integer values of $p \geq -2$ this representation can
 be given by a tensor $\cO_{ABC_1\ldots C_p}$
which is (generalised) antisymmetric on all $n=p+2$ indices. It is
not difficult to check that all such tensors do indeed have the same number of
components for $p=0,1,2,\ldots$. We now define the quasi-tensor $\cO_{AB}[p]$ to be the object with components

 \bea
 \cO_{33}[p]&=&\cO_{3333\ldots} \\
 \cO_{\a 3}[p]&=& \cO_{\a 333\ldots} \\
 \cO_{\a\b}[p]&=& \cO_{\a\b 33\ldots}
 \la{35}
 \eea

All of the other components of $\cO$ vanish by antisymmetry of the
even indices. If we take $\gs \gl (2|1)$ to act from the left, we find
its action on these tensors to be
\bea
 \d\cO_{33}[p]&=&(p+2)A_3{}^{\a}\cO_{\a 3}[p]+ (p+2)A_3{}^3\cO_{}[p] \\
 \d\cO_{\a 3}[p]&=&\hat A_{\a}{}^{\b}\cO_{\b 3}[p]+
 (p+{3\over2})A_3{}^3\cO_{\a 3}[p]
 +A_{\a}{}^3\cO_{33}[p]+ (p+1)A_3{}^{\b}\cO_{\a\b}[p]\\
 \d\cO_{\a\b}[p]&=&(p+1)A_3{}^3\cO_{\a\b}[p] -2A_{[\a}{}^3\cO_{\b]3}[p]
 \la{36}
 \eea
where $A_A{}^B\in\gs\gl(2|1)$ and $\hat A_{\a}{}^{\b}$ is
traceless. Now this formula makes sense for arbitrary values of $p$
(even complex) and it is straightforward to check that one still
has a representation of the algebra. In the context of unitary
representations of the superconformal group we are interested in $p$
real and $p \geq -1$ or $p=-2$. For $p=-1$ the representation becomes reducible,
although not completely reducible. The components $(\cO_{33}[-1], \cO_{\a 3}[-1])$
transform under a subrepresentation, while the quotient
representation in this case is just a singlet. For the case $p=-2$ we
see that the component $O_{33}[-2]$ is invariant and we obtain the
trivial representation. These two cases are atypical representations whereas
for $p > -1$ the representations are typical.

\subsection{Representations on various superspaces}

We are now in a position to give any representation as a
superfield on different superspaces. Note that
unitary irreducible representations can not be carried by holomorphic fields on super
twistor space. On super twistor space~(\ref{stwistor}) we would
obtain fields transforming under the representation of the
semisimple part of the Levi subalgebra $\gl_S=\gs\gl(3|N)$ with
the following Dynkin diagram \be
\begin{picture}(200,30)(-10,-10)
\put(0,0){\makebox[0pt][l]{$\ominus\hspace{1.5em}
    \bt \hspace{1.5em}
\bt\hspace{1.5em}\cdots
\hspace{1.5em}\bt\hspace{1.5em}\bt\hspace{1.5em}\ominus\hspace{1.5em}
\bt$}} \put(0.5,0){\makebox[0pt][l]{\rule[.5ex]{5.1em}{.1ex}
    $\hspace{4.2em}$
\rule[.5ex]{6.9em}{.1ex}}}

\put(0,10){\tiny $n_2 \hspace {2em}n_3
 \hspace{2em}n_4
    \hspace{6em}n_{N}\hspace{1.5em}n_{N+1}\hspace{1.5em}n_{N+2}
    \hspace{1.5em}n_{N+3}$}
\end {picture}
\ee
This diagram is not in the distinguished basis as it contains two odd
nodes. If we rearrange this in terms of the distinguished set of simple roots
of $\gs\gl(3|N)$ we get
\be
\ba{c}
\begin{picture}(200,30)(-10,-10)
\put(0,0){\makebox[0pt][l]{$\bt\hspace{1.5em}\bt\hspace{1.5em}
    \ominus\hspace{1.5em}
\bt\hspace{1.5em}\bt\hspace{1.5em}\cdots
\hspace{1.5em}\bt\hspace{1.5em}\bt$} \rule[.5ex]{9em}{.1ex} $\hspace{4.3em}$
\rule[.5ex]{2.2em}{.1ex}
 }
\put(0,10){\tiny $n' \hspace {2em}n_{N+3}
 \hspace{1.5em}n''\hspace{2em}n_{3}\hspace{2em}n_{4}
    \hspace{6em}n_{N}\hspace{2em}n_{N+1}$}
\end {picture}\\
n'=-n_2-n_{N+2}-m_1 \qquad n''=n_{N+3}-n_{N+2}-m_1
\ea
\ee
 where $m_1$ is defined in~(\ref{m}). We see that the
 coefficient $n'$ is negative and therefore the representation is not
 finite dimensional (recall from the paragraph following~(\ref{lambda})
 that a representation is finite dimensional if and only if all the distinguished even
 coefficients are positive). We therefore
 exclude this space from our analysis. We also exclude dual projective
 twistor space that
 has one cross
 on the right hand node of the Dynkin diagram for $SL(4|N)$, and
 ambitwistor space,
 which has crosses on both extremal nodes, for the same reason. Note
 that the problem with
 these spaces is the
 same as in the bosonic case using twistor space
 (section~\ref{twistor}). One must presumably use higher cohomology to
 treat
 these cases as with twistor space.

 We can, however give any unitary representation of
 the superconformal group as a holomorphic superfield on any other superflag
 space. These superfields will sometimes have to satisfy constraints
 to make them irreducible representations. However, on analytic
 superspaces the fields require no constraints as shown in section~\ref{sect:red}.

 For the reasons just given, we will mainly be interested in
 analytic superspaces. A unitary representation on $(N,p,q)$ analytic
 superspace has Dynkin diagram
 \be
\begin{picture}(220,20)(-10,0)
\put(0,0){\makebox[0pt][l]{$\bt\hspace{1.5em}\ominus\hspace{1.5em}\bt\hspace{1.5em}
\times\hspace{1.5em}\bt
\hspace{1.5em}\times\hspace{1.5em}\bt\hspace{1.5em}\ominus
\hspace{1.5em} \bt$} \rule[.5ex]{4.5em}{.1ex} $\hspace{.2em}
\cdots \hspace{1em} \cdots \hspace{1em} \cdots \hspace{1em}\cdots
\hspace{.4em}$ \rule[.5ex]{5.0em}{.1ex}
 }
 \put(0,10){\tiny $n_1 \hspace{2em} n_2 \hspace{2em} n_3
 \hspace{2em}\hspace{10.5em} n_{N+1} \hspace{.5em}n_{N+2} \hspace{.5em}
 n_{N+3}$} \end {picture}
\ee
 The fields transform under the supergroups $\gs\gl(2|p)$ and
 $\gs\gl(2|q)$ under the representation given by the first $p+1$
 nodes and the last $q+1$ nodes. But as mentioned, the unitarity conditions
~(\ref{bounds}) tell us that these are either trivial representations
 or contravariant
 representations (compare with~(\ref{condition}).) In other words,
 on analytic superspaces the unitary representations are
 characterised by the fields with downstairs supergroup indices.
 In fact this can be made into a general statement about any
 superspace (except the twistor spaces mentioned above): the
 unitary representations are precisely those fields with
 downstairs superindices.

\subsection{$N=2$ analytic superspace}

We now look at some specific examples on complex $N=2$ analytic
superspace. This has the following Dynkin diagram
 \be
\begin{picture}(30,10)(0,0) \put(0,0){\makebox[0pt][l]{$\bt
\hspace {2em}
 \ominus \hspace{2em}\times \hspace{2em}\ominus \hspace{2em}\bt$} \rule[.5ex]{11.6em}{.1ex} }
 \end{picture}
\ee
 with corresponding parabolic subgroup consisting of matrices of the
 form
\be
\left( \ba{ll} a^A{}_B & 0 \\ c_{A'B} & d_{A'}{}^{B'} \ea \right) \label{par}
\ee
where each entry is a $(2|1) \times (2|1)$ matrix. The Levi subalgebra
(under which our fields transform) is $\gs \gl(2|1) \oplus \gs \gl(2|1)\oplus
\com$ (corresponding to the block diagonals), where the first $\gs
\gl(2|1)$ subalgebra  is carried by un-primed
indices, and the second by primed
indices.

 One needs two coordinate charts $U$ and $U'$, with coordinates
 $X$ and $X'$, to cover the whole of complex
analytic superspace. Coset representatives for each of these
charts can be given as follows
 \bea
 s_1(X)&=& \left( \ba{c|cc|c} 1_2&0&\l&x\\
 \hline                     0&1&y&\pi\\
                            0&0&1&0\\
 \hline                     0&0&0&1  \ea \right) \label{s1}\\
 s_2(X')&=& \left( \ba{c|cc|c} 1_2&\l'&0&x'\\
 \hline                     0&y'&1&\pi'\\
                            0&1&0&0\\
 \hline                     0&0&0&1_2  \ea \right).
 \eea
 These are related by a
superconformal transformation, i.e.
 \be
 s_2(X')=s_1(X')K, \qquad K=\left(
 \ba{c|cc|c}1_2&0&0&0\\ \hline 0&0&1&0\\0&1&0&0\\
 \hline 0&0&0&1_2
 \ea \right).
 \ee
 By performing a compensating isotropy group transformation, $s_2(X')=h
 s_1(X(X')))$ we
 can relate the two sets of coordinates on $U \cap U'$:
 \be
 \left( \ba{c|cc|c} 1_2&\l'&0&x'\\
 \hline                     0&y'&1&\pi'\\
                            0&1&0&0\\
 \hline                     0&0&0&1_2  \ea \right)
 =
 \left(
 \ba{c|cc|c}1_2&\l'&0&0\\ \hline 0&y'&0&0\\
 0&1&-{1 \over y'}&-{\pi' \over y'}\\
 \hline 0&0&0&1_2
 \ea \right)\left( \ba{c|cc|c} 1_2&0&-{ \l' \over y'}& x'-{\l' \pi' \over y'}\\
 \hline                     0&1&{1 \over y'}&{\pi' \over y'}\\
                            0&0&1&0\\
 \hline                     0&0&0&1_2  \ea \right) \label{coords}
 \ee
giving
\be
\ba{rcl}
x&=&x'-{\l' \pi' \over y'}\\
\l&=&-{\l' \over y'}\\
\pi&=&{\pi' \over y'}\\
y&=&{1 \over y'} \ea \label{X'}
 \ee
 Requiring our fields to be
holomorphic on both patches puts restrictions on the fields, which
are equivalent to the constraints on Minkowski space.

We use the formalism of induced
representations~(\ref{F(hu)=}-\ref{g.F}), representations of parabolic
subgroups, and
superindices outlined above to find the transformation
of fields. Define $f(X)=F(s_1(X))$ and $f'(X')=F(s_2(X'))$, then
under the superconformal transformation mapping $X \mapsto X'$ we
have
 \be f(X) \mapsto f'(X')=R(h)f(X)\label{f'(X')}
 \ee
 where $h$ is given in~(\ref{coords}). However only the Levi part of $h$
 \be l=\left(
 \ba{c|cc|c}1_2&\l'&0&0\\ \hline 0&y'&0&0\\
 0&0&-{1 \over y'}&-{\pi' \over y'}\\
 \hline 0&0&0&1_2
 \ea \right)
 =\left(
 \ba{c|cc|c}1_2&-{\l \over y}&0&0\\ \hline 0&{1 \over y}&0&0\\
 0&0&- y&-\pi\\
 \hline 0&0&0&1_2
 \ea \right)
\ee
 acts non-trivially. Comparing with \eq{par} we find
 \be
a^A{}_B= \left( \ba{c|c}  1_2& -{\l \over y}\\
\hline            0            &{1 \over y} \ea \right) \qquad
d_{A'}{}^{B'}= \left( \ba{c|c} -y& -\pi\\
\hline            0            &  1_2 \ea \right).
 \ee
 For a representation with Dynkin coefficients $n_i$ we have the
Dynkin diagram
 \be
 \begin{picture}(90,10)(0,0) \put(0,0){\makebox[0pt][l]{$\bt
\hspace {2em}
 \ominus \hspace{2em}\times \hspace{2em}\ominus \hspace{2em}\bt$}
 \rule[.5ex]{11.6em}{.1ex} }
 \put(0,10){\tiny $n_1 \hspace{2.5em} n_2 \hspace {3.3em}n_3
 \hspace{3.3em}n_4 \hspace{3em}n_5$}
 \end{picture}
\ee
  find that
  \be
  R(h)=y^{-Q}
R^{(n_1,n_2)}(a^A{}_B)R^{(n_4,n_5)}(d_{A'}{}^{B'})\label{R(h)}
 \ee
 where $R^{(n_1,n_2)}$ and $R^{(n_4, n_5)}$ are the
 representations
 of $SL(2|1)$ with
 Dynkin diagrams
 \be
 \begin{picture}(30,10)(0,0) \put(0,0){\makebox[0pt][l]{$\bt
\hspace {2em}
 \ominus $} \rule[.5ex]{2.87em}{.1ex} }
 \put(0,10){\tiny $n_1 \hspace{2.5em} n_2 $}
 \end{picture}
 \hspace{10em}
 \begin{picture}(30,10)(0,0) \put(0,0){\makebox[0pt][l]{$\bt
\hspace {2em}
 \ominus $} \rule[.5ex]{2.87em}{.1ex} }
 \put(0,10){\tiny $n_5 \hspace{2.5em} n_4 $}
 \end{picture}
\ee
 respectively, extended to a representation of $GL(2|2)$ in the natural
 way. By considering the corresponding Cartan
 generators acting on the highest weight state one can show that
 \be
 Q=-n_1+n_2+n_3+n_4-n_5 = L- J_1 - J_2 \label{Q}
 \ee
 in the case that the supergroup representations are contravariant,
 and is generated by the Cartan matrix diag$(-1,0|0,0|0,1)$.

 To find the representations $R^{(n_1,n_2)}$ and $R^{(n_4, n_5)}$ one
 first converts the Dynkin diagrams into Young Tableaux. From the
 unitary bounds~(\ref{bounds}) and the condition~(\ref{condition}) we
 know that we will always obtain trivial or contravariant representations of
 both the $SL(2|1)$ subgroups. There is some choice as to how these
 transform (see discussion in the previous subsection.) We will
 choose our fields so as to transform consistently with $dX^{AA'}$
 and
 $\partial/\partial X^{AA'}.$ Under the transformation
~(\ref{X'}) we obtain
 \bea
 \left( \ba{cc} d\l' & dy'\\ ds' & d\pi' \ea
 \right)&=&
 \left( \ba{cc} 1 & {-\l \over y}\\ 0 & {1 \over y} \ea
 \right)
\left( \ba{cc} d\l & dy\\ ds & d\pi \ea
 \right)
 \left( \ba{cc} -{1 \over y} & -{\pi \over y}\\ 0 & 1 \ea
 \right)
 \\
 \left( \ba{cc} \del_{\l'} & \del_{y'}\\ \del_{s'} & \del_{\pi'} \ea
 \right)&=&
 \left( \ba{cc} -y & -\pi\\ 0 & 1 \ea
 \right)
\left( \ba{cc} \del_{\l} & \del_{y}\\ \del_{s} & \del_{\pi} \ea
 \right)
 \left( \ba{cc} 1 & -\l\\ 0 & y \ea
 \right)
 \eea
which we can rewrite
 \bea
 dX'^{BB'} &=&  a^B{}_A dX^{AA'}(d^{-1})_{A'}{}^{B'}\\
 \del'_{A'A} &=&  d_{A'}{}^{B'}\del_{B'B} \tilde{a}^B{}_A.
 \eea
 where we define
 \be
 \tilde{a}^B{}_A = (a^{-1})^B{}_A (-1)^{(A+B)}.
 \ee

We thus define tensors to transform as follows
 \be \ba{rclcrcl}
 W^A &\mapsto& a^A{}_B W^B&&W^{A'} &\mapsto&
 W^{B'}(d^{-1})_{B'}{}^{A'}\\
 W_{A}&\mapsto& W_B \tilde{a}^B{}_A&&W_{A'} &\mapsto&
 d_{A'}{}^{B'}W_{B'}\label{tensor}
\ea \ee
 and we see that the combinations $W^AV_A$ and $W^{A'}V_{A'}$ are
 scalars, whereas $V_AW^A$ and $V_{A'}W^{A'}$ are not.

 In section~\ref{sec:an} we gave a few examples of $N=2$
 supermultiplets as fields on
 analytic superspace. We look at these examples now in more
 detail. The simplest example of a field on analytic superspace is
 that of the hypermultiplet. The Dynkin diagram
 for this representation on analytic superspace
 \be
\begin{picture}(30,10)(0,0) \put(0,0){\makebox[0pt][l]{$\bt
\hspace {2em}
 \ominus \hspace{2em}\times \hspace{2em}\ominus \hspace{2em}\bt$} \rule[.5ex]{11.6em}{.1ex} }
 \put(0,10){\tiny 0\hspace{3.3em} 0 \hspace{3.3em} 1 \hspace {3.5em}0
 \hspace{3.3em} 0}
 \end{picture}
 \ee
 tells us that it has no superindices (since $n_1=n_2=n_4=n_5=0$)
 and that $Q=1$ \eq{Q}. Thus from~(\ref{f'(X')},\ref{R(h)}) it is a field
 $W$ which transforms as
 \be
 W(X) \mapsto W'(X')={1 \over y} W(X) \label{hyp}
 \ee
 under the transformation $X \mapsto X'$~(\ref{X'}). Demanding
 that $W(X)$ and $W'(X')$ are both holomorphic in their variables
 leaves us with the component fields of the hypermultiplet
 \be
 \ba{rcl}
 W(x,\l,\pi,y)& =& \vf_1(x) + y \vf_2(x) + \l^{\a} \psi_{\a}(x)\\&
+& \pi^{\adt}\chi_{\adt}(x) -\l^{\a}\pi^{\adt} \partial _{\a \adt}
\vf_2\\
W'(x',\l',\pi',y')& =& \vf_2(x') + y' \vf_1(x') - \l'^{\a}
\psi_{\a}(x')\\& +& \pi'^{\adt}\chi_{\adt}(x')
-\l'^{\a}\pi'^{\adt}
\partial _{\a \adt} \vf_1
 \ea
 \ee
 and with the component fields $(\vf_1,\vf_2,\psi,\chi)$ all
 satisfying their equations of motion \cite{Eden:2000qp}.

 Next consider the $N=2$ on-shell Maxwell multiplet, which is usually
 given as a chiral field in super Minkowski space. It has dilation
 weight 1, R-charge -1 and all other quantum numbers are 0 and thus
 has the following Dynkin diagram as a field on analytic superspace
\be
 \begin{picture}(180,20)
\put(20,0){\makebox[0pt][l]{$\bt\hspace{3em}\ominus\hspace{3em}\times
    \hspace{2.7em}
\ominus \hspace{3em}\bt$} \rule[.5ex]{15.5em}{.1ex}

 }
\put(20,10){\tiny  0
\hspace{4.5em} 1 \hspace{5em} 0 \hspace{4.5em} 0 \hspace{4.5em}
0}
\end {picture}
\ee
 We can read off from the Dynkin diagram that the field transforms
 trivially under the second $SL(2|1)$ factor, but under the first
 $SL(2|1)$ factor it transforms in the representation
 \be
 \begin{picture}(30,10)(0,0) \put(0,0){\makebox[0pt][l]{$\bt
\hspace {2em}
 \ominus $} \rule[.5ex]{2.87em}{.1ex} }
 \put(0,10){\tiny $0 \hspace{3.5em} 1 $}
 \end{picture}
\ee
 and $Q=1$. This corresponds to a field $W_A$ with one downstairs, unprimed
superindex, transforming as
 \be
 W'_A = {1 \over y} W_B \tilde{a}^B{}_A\label{Max}
 \ee
 which, if we let $W_A=(W_{\a},W)$ becomes
 \be
 \left( \ba{c} W'_\a\\W' \ea \right)=\left( \ba{c}
 W_{\a}/y\\-W_{\b} \l^{\b}/y +W \ea \right)
 \ee
 Once again, demanding that both $W_A$ and $W'_A$ are holomorphic
 in their respective variables leaves the correct components
 \be
 \ba{rcl}
 W_{\a}&=&\r_{1\a}+y \r_{2\a}+\l^{\b}F_{\a\b} -
 \pi^{\adt}\del_{\a\adt}C-\l^{\b}\pi^{\bdt}\del_{\b\bdt}\r_{2\a}\\
 W&=&C-\l^{\a}\r_{2\a}
 \ea
 \ee
 with the components $(\r_{1\a},\r_{2\a}, F_{\a\b},C)$
 all satisfying their equations of motion.

 The conjugate representation is similar. It is usually given as
 an antichiral field on super Minkowski space and has conjugate Dynkin
 diagram to~(\ref{chiral}) on analytic superspace. It is therefore given
 as a field $W_{A'}$ with $Q=1 $, transforming as $W_{A'} \mapsto
 W'_{A'}$ with
 \be
 W'_{A'}={1 \over y}d_{A'}{}^{B'}W_{B'}.
 \ee
 Its components are the on-shell fields
 $(\s_{1\adt},\s_{2\adt},G_{\adt\bdt},D)$.

As a final example, consider the $N=2$ superconformal stress-energy multiplet. On super Minkowski space it is a scalar superfield $T$ satisfying
\be
D_{\a i} D_j^{\a} T =0.
\ee
It has dilation weight 2, and all other quantum numbers are 0. On
analytic superspace it has the Dynkin diagram
\be
 \begin{picture}(170,20)(10,0)
\put(20,0){\makebox[0pt][l]{$\bt\hspace{3em}\ominus\hspace{3em}\times
    \hspace{2.7em}
\ominus \hspace{3em}\bt$} \rule[.5ex]{15.5em}{.1ex}

 }
\put(20,10){\tiny  0 \hspace{4.5em} 1 \hspace{5em} 0
\hspace{4.5em} 1 \hspace{4.5em} 0}
\end {picture}
\ee
 and thus it is given by the superfield $T_{A'A}$ with $Q=2$. It
 thus transforms under $X \mapsto X'$ as $T_{AA'} \mapsto
 T'_{AA'}$ where
  \be
  T'_{A'A} =  {1 \over y^2}
 d_{A'}{}^{B'}T_{B'B}\tilde{a}^B{}_A. \label{stress}
 \ee
 Then from demanding that both $T_{A'A}$ and $T'_{A'A}$ are
 holomorphic in their respective variables leaves us with the
 correct components of the stress energy multiplet, satisfying
 the correct equations.

 As mentioned previously this representation can also be realised explicitly in two
 different ways on analytic superspace: firstly by multiplying
 a Maxwell field and its conjugate together \eq{WW}
 and secondly by multiplying two hypermultiplet fields together
with a derivative \eq{WdW}.

 \subsection{$(4,2,2)$ analytic superspace}

 A highest weight representation on $(4,2,2)$ analytic superspace has
 Dynkin diagram
 \be
 \begin{picture}(192,10)(0,0) \put(0,0){\makebox[0pt][l]{$\bt
 \hspace {2em}
 \ominus \hspace{2em}\bt\hspace{2em}\times\hspace{2em}\bt
 \hspace{2em}\ominus \hspace{2em}\bt$}
 \rule[.5ex]{17.1em}{.1ex} }
 \put(0,10){\tiny $n_1 \hspace{2.5em} n_2 \hspace {3em}n_3
 \hspace{3.5em}n_4\hspace{3em}n_5\hspace{3em}n_6 \hspace{3em}n_7$}
 \end{picture}
 \ee
 This has Levi subgroup $S(GL(2|2)\xz GL(2|2))$ and has coordinates
 $X^{AA'}$ given in equation  \eq{N=4}. Primed and unprimed capital
 indices carry representations of the two $GL(2|2)$
 subsupergroups. Using the method of induced
 representations~(\ref{F(hu)=}-\ref{g.F}), we
 consider equivariant maps $F:SL(4|4) \rightarrow V$ where $V$ is the
 particular
 representation space for the parabolic subgroup given by the Dynkin
 diagram.  So we have maps such that
 $F(hu)=R(h)F(u)$ where $u \in SL(4|4)$ and
 \be
 h=\left( \ba{ll} a^A{}_B & 0 \\ c_{A'B} & d_{A'}{}^{B'} \ea \right)
 \ee
 with each entry a $(2|2)$ matrix. The representation $R$ is given by
 \be
 R(h)=|a|^{-Q} R^{(n_1,n_2,n_3)}(a) R^{(n_5,n_6,n_7)}(d)
 \ee
 where $R^{(n_1,n_2,n_3)}(a)$ and $R^{(n_5,n_6,n_7)}(d)$ are tensor
 representations of $SL(2|2)$ extended to $GL(2|2)$ given by the
 Dynkin labels, $|a|$ is the superdeterminant of $a$ and
 \be
 Q= -n_1 + n_2 +n_3 + n_4 +n_5 +n_6 -n_7 = L-J_1-J_2.
 \ee

 The superalgebra $\gs \gl(4|4)$ is not simple, as it contains the identity
 matrix and thus has a non-zero centre. In order to obtain a simple
 superalgebra, one has to exclude
 this to obtain $\gp \gs \gl(4|4)$ which is simple. Representations of
 the simple super Lie group
 $PSL(4|4)$ are representations of $SL(4|4)$ for which the centre does
 not act. From \eq{gen} we see that the centre is generated by
 $\hat{R}$ and so representations of $PSL(4|4)$ are representations of
 $SL(4|4)$ with $R=0$. From \eq{ni} we obtain
 \be R={1 \over
 2}(n_1-2n_2-n_3+n_5+2n_6-n_7)
 \ee
 and by considering Young Tableaux one can show that the number of
 unprimed indices is $n_3 + 2n_2-n_1$ and the number of primed indices
 is $n_5 + 2n_6 -n_7$ Thus the condition $R=0$ tells us that for
 representations of $PSL(4|4)$ the
 number of primed indices equals the number of unprimed
 indices.

 The only massless multiplet with $R=0$ is the Maxwell multiplet
 given on (4,2,2) analytic superspace by a field $W$ with charge $Q=1$
 and with no superindices. We can obtain
 other representations of $PSL(4|4)$ by taking copies of the Maxwell
 field and applying derivative operators $\del_{A'A}$ to them.
 For example, consider the representation with internal Dynkin
 labels $[1,0,1]$, dilation weight $L=2$ and 0 spin. On $(4,2,2)$
 space this is given by a field $W_{A'A}$ with charge $Q=2$. Since the
 derivative operator $\del_{A'A}$ has zero charge\footnote{This is
 true as long as we only symmetrise the indices as dictated by Young
 tableau with $m_1=0$.}, we need two copies
 of the Maxwell field, $W^{(1)}$ and  $W^{(2)}$, and one derivative
 operator to obtain this
 field. In order for the field to be primary we must take the
 combination \eq{comb} $W_{A'A}=W^{(1)}\del_{A'A}W^{(2)} -
 W^{(2)}\del_{A'A}W^{(1)}$.

 In fact, one can obtain all unitary irreducible representations of $PSL(4|4)$ (with
 integer Dynkin labels) in this
 manner. Given a representation with super Dynkin labels $(n_1, \dots, n_7)$
 we need $Q$ $W$'s and $n_3 + 2n_2-n_1=n_5 + 2n_6 -n_7$ $\del$'s. We may
 need to use higher order derivatives such as $\del_{A'A}\del_{B'B}$,
 in which case the
 primed and unprimed super indices will have the same symmetry
 properties since the $\del$'s commute. However it is not hard to convince
 oneself that for any
 given representation, we can take a particular combination of
 $\del$'s acting on $W$'s to obtain that representation. There are
 many different ways of doing this, we give just one way. We have
 $Q$ copies of the Maxwell multiplet $W_i:(i=1\dots Q)$.
 If $n_1<n_7$ we can take the following combination
 \be
 \del^{n_1+2}W_1 \prod_{i} \del^2 W_i
 \prod_{j}\del W_j \prod_k W_k
 \ee
 where $i,j,k$ are
 \be \ba{rcl}
 n_3<n_5 &\quad \Rightarrow \quad & \left\{ \ba{l}i=(2,\dots, n_2-n_1)\\
 j=(n_2-n_1+1, \dots, n_3+n_2-n_1)\\
 k=(n_3 +n_2-n_1+1, \dots, Q) \ea \right. \\ \\
 n_5 < n_3 &\quad \Rightarrow \quad & \left\{ \ba{l}i=(2,\dots,\lfloor
 n_6-{n_7
 +n_1 \over 2}\rfloor) \\
 j=(\lfloor n_6-{n_7
 +n_1 \over 2}\rfloor), \dots,\lceil n_5+n_6-{n_7+n_1 \over 2}\rceil\\
 k=(\lceil n_5+n_6-
 {n_7+n_1 \over 2}+1\rceil, \dots, Q)\ea \right.
 \ea \ee
 Here $\lfloor x \rfloor$ ($\lceil x\rceil)$ denote the nearest integers
 less than (greater
 than) or equal to $x$.
 Whenever more than one $\del$ acts on a $W_i$ we take the
 symmetric combination on both indices. We then project onto the
 representation of the two $SL(2|2)$ subgroups with the following
 Young tableaux
 \be \setlength{\unitlength}{.2mm}
 \begin{picture}(210,100)(0,60)
 \put(60,140){\framebox(100,20){}}\put(100,147){$n_1$}
 \put(20,60){\framebox(20,100){}}\put(42,70){$\left. \right\}n_3$}
 \put(20,90){\framebox(40,70){}}\put(-65,117){$n_2-n_1\left\{ \ba{c} \\ \\ \\ \ea \right.$}
 \put (17,45){(unprimed Indices)}
 \end{picture}
 \qquad
 \begin{picture}(200,100)(0,60)
 \put(60,140){\framebox(100,20){}}\put(100,147){$n_7$}
 \put(20,60){\framebox(20,100){}}\put(42,70){$\left. \right\}n_5$}
 \put(20,90){\framebox(40,70){}}\put(-65,117){$n_6-n_7\left\{ \ba{c} \\ \\ \\ \ea \right.$}
 \put (17,45){(primed Indices)}\end{picture}
 \ee
 One will have to add other terms in order to make the operator
 superconformally covariant.

 Note that although we can obtain all unitary irreducible representations of
 $PSL(4|4)$ with integer Dynkin coefficients in this manner many of
 the operators of interest - in $N=4$ super Yang-Mills for
 instance - have anomalous dimensions, which lead to non-integer odd
 Dynkin coefficients. Such operators can not be given by
 multiplying together $W$'s in the above fashion, but they can be
 represented in analytic
 superspace as abstract `quasi-tensor' superfields (section~\ref{quasi}).

 \section{Oscillator method} \la{oscillator}

G\"unaydin's oscillator method has been used to obtain the unitary
irreducible representations of the conformal group $SU(2,2)$
explicitly~\cite{Bars:1983ep,Gunaydin:1985fk,Gunaydin:1998sw,Gunaydin:1998jc,Gunaydin:1999jb}.
These representations can be extended naturally to give
representations (no longer unitary) of the complex conformal group
$SL(4)$ and one can relate these representations to those acting
on fields on complex Minkowski space. In this way we will be able
to see that the representations acting on our fields are also
unitary. Similarly one can also use the oscillator method to
construct unitary representations of the superconformal group
$SU(2,2|N)$~\cite{Bars:1983ep,Gunaydin:1985fk,Gunaydin:1998sw,Gunaydin:1998jc,Gunaydin:1999jb}.
If one tries to relate these representations to fields in an
analogous way to the bosonic case one naturally obtains fields on
analytic superspace.

\subsection{The bosonic case}

We here show the oscillator method of obtaining representations of
the complex conformal group $SL(4)$ which correspond to unitary
representations of the real conformal group $SU(2,2)$,
following~\cite{Gunaydin:1998sw,Gunaydin:1998jc,Gunaydin:1999jb}.
The lie algebra $\gs \gl(4)$ is the set of 4 $\times$ 4 traceless
matrices:
 \be
 \gs \gl(4)=\left\{\left( \ba{cc} A^{\a}{}_{\b} & B^{\a \bdt}\\
                     C_{\adt \b}   & D_{\adt}{}^{\bdt}
       \ea \right) \right\}\label{SL4}
\ee
 In the following we will refer to the Levi decomposition of $\gs
 \gl(4)$ corresponding to Minkowski space, i.e. $\gs\gl(4)=\gn^-
 \oplus \gl \oplus \gn^+$ where $\gl$ is the set of block diagonal
 matrices (matrices in the previous equation such
 that $B=C=0$), $\gn^-$ and $\gn^+$ are the upper-right and
 lower-left blocks respectively.
 One takes $P$ copies of oscillator operators $a^{\a}(K)=
 (a_{\a}(K))^{\dagger},\ b^{\adt}=(b_{\adt})^{\dagger}$
 transforming under the
two $SL(2)$ subgroups corresponding to the matrices $A$ and $D$.
Here $K, L = 1, \dots, P$ label different generations of
oscillators. The oscillators satisfy the following commutation
relations:
 \be [a_{\a}(K),a^{\b}(L)] = \d_{\a}^{\b} \d_{L
 K} \qquad
 [ b_{\adt}(K),b^{\bdt}(L)] = \d^{\bdt}_{\adt}
\d_{L K} \label{com}
 \ee
 with all other commutation relations zero. Oscillators with an upper (lower)
 index are creation (annihilation) operators.
 The `vacuum' $|0\rangle$ is defined
 by
 \be a_{\a}|0\rangle = b_{\adt}|0\rangle=0.  \ee
 It is helpful to define 4-spinors
 $\psi^{\ua}$, $\bar{\psi}_{\ua}$ as follows
 \be
 {\psi}^{\ua}=(a^{\a},  b_{\adt})^T
  \qquad
\bar{\psi}_{\ua}=(\psi^{\dagger})^{\ub}J^{\ua}_{\ub}=(
 a_{\a},-b^{\adt}).
  \la{psi}\ee
 where  $J=\mbox{diag}(1,1,-1,-1)$
 so that they satisfy the commutation relation
 \be
  [{\psi}^{\ub}(L),\bar{\psi}_{\ua}(K)]=-\d_{\ua}^{\ub}\d_{KL}.
  \ee
 If we call the matrix in~(\ref{SL4}) $M^{\ua}{}_{\ub}$ and define
 \be
 \widehat M:= - \sum_{K=1}^P \bar{{\psi}}_{\ua}(K)M^{\ua}{}_{\ub}{\psi}^{\ub}(K)
 \ee
 we find
 \be
 [\widehat M , \widehat M' ] = \widehat{[M,M']}.
 \ee
 So $\widehat M$ is an infinite dimensional representation of
 $\gs \gl(4)$ acting on the Fock space of the oscillators.

 The dilation weight $L$ is generated by
 \be
 \widehat{L}= {1 \over 2} ( a_{\a} \cdot a^{\a} + b^{\adt} \cdot b_{\adt}) = {1
 \over 2}(N_a + N_b +2P)\label{dil}
 \ee
 where the dot corresponds to summation over the $P$ different
 oscillator generations, and where $N_a=a^{\a} \cdot a_{\a}$,
 $N_b=b^{\adt} \cdot b_{\adt}$ are
 the bosonic number operators.
 The irreducible representations of $\gs \gl(4)$ are obtained by
 constructing an irreducible representation, $\langle \O|$, of the
 parabolic subalgebra $\gl \oplus \gn^+$ in the
 Fock space of the oscillators. As mentioned in the appendix, irreducible
 representations of parabolic subgroups are acted on trivially by the
 nilpotent part $\gn^+$, therefore we insist that $\langle \O|$ is annihilated by the
 generators which correspond to the generators $\gn^+$
 \be
 \langle \O| a^{\a} \cdot b^{\adt}  =0.
 \ee
 Applying
 the operators $a_{\a} \cdot b_{\adt}$ (which correspond to
 $\gn^-$)
 repeatedly on $\langle \O|$
 one generates the infinite set of states
 \be
 \langle \O|, \qquad \langle \O|a_{\a} \cdot b_{\adt},\qquad
 \langle \O| a_{\a} \cdot
 b_{\adt}a_{\b} \cdot b_{\bdt}, \qquad \dots
 \ee
 which form a basis for the irreducible representation of $SL(4)$.
 Similarly a set of states are obtained from the conjugate representation
 $|\O\rangle$, which is annihilated by $a_{\a} \cdot b_{\adt}$
 \be
 |\O\rangle, \qquad a^{\a} \cdot b^{\adt}|\O\rangle ,\qquad
 a^{\a} \cdot
 a^{\adt}a^{\b} \cdot b^{\bdt} |\O\rangle , \qquad \dots.
 \ee

 The equivalence between this formulation of the
 representations, and field representations on complex Minkowski
 space is straightforward.
 The state $|\Phi\rangle$ corresponds to the field $\Phi(x)$ given
 by
 \be
 \Phi(x^{\a \adt}) = \langle \O | \widehat{s}(x) |\Phi\rangle
 \label{relation}
 \ee
 where $\widehat{s}(x)=\mbox{exp}(x^{\a \adt} a_{\a} \cdot
 b_{\adt})$ corresponds to the coset representative of $\gs\gl(4)$
 for complex Minkowski space~(\ref{coset}). The field $\Phi(x)$
 therefore inherits the index structure of $\langle \O|$. Under an
 $\gs \gl(4)$ transformation $|\Phi\rangle \mapsto \widehat{M}
 |\Phi\rangle$ the field transforms as
 \be
 \Phi(x) \mapsto \langle \O| \widehat{s}(x) \widehat{M} |\Phi\rangle =
 \langle \O| \widehat{h}(x,g)\widehat{s}(x')|\Phi\rangle =
 R(h(x,g))\Phi(x')
 \ee
 exactly as an induced representation should~(\ref{g.F}).
 Explicitly the
 correspondence can be written in terms of the Taylor series for
 $\Phi(x)$ as
 \be
 |\Phi\rangle = \Phi(0) |\O\rangle + \del_{\a \adt}\Phi(0) a^{\a}\cdot
 b^{\adt}|\O\rangle + \dots.
 \ee

 For example, if we take $P=1$ so there is only one generation of
 oscillators, then the only possible forms $|\O\rangle$ can have are
 \be
 \langle0| a_{\a_1}\dots a_{\a_{2J_1}}
 \ee
 and
 \be
 \langle0| b_{\adt_1}\dots b_{\adt_{2J_2}} .
 \ee
 Using the expressions for the dilation weight~(\ref{dil})
 and~(\ref{dy}) we find
 that these have Dynkin coefficients
 \be
\begin{picture}(70,10)(0,0) \put(0,0){\makebox[0pt][l]{$\bt
\hspace {2em}\xz \hspace{2em}\bt$} \rule[.5ex]{5.5em}{.1ex} }
 \put(0,10){\tiny $2J_1 \hspace{.2em} -(2J_1+1) \hspace {1.3em}0$}
 \end{picture}
\qquad \qquad \qquad
\begin{picture}(70,10)(0,0) \put(0,0){\makebox[0pt][l]{$\bt
\hspace {2em}
 \xz \hspace{2em}\bt$} \rule[.5ex]{5.5em}{.1ex} }
 \put(0,10){\tiny $0 \hspace{.4em} -(2J_2 +1) \hspace {1.5em}2J_2$}
 \end{picture}.
\ee
 These correspond to massless fields on Minkowski space.
 For example, take $\langle\O | = \langle 0 |$ then we obtain the
 representation corresponding to a massless scalar. Using the
 relation~(\ref{relation}) notice that
 \be
 \del_{\a \adt} \del_{\b \bdt}
 \Phi(x)=\langle0|a_{\a}b_{\adt}a_{\b}b_{\bdt}
 \widehat{s}(x)|\Phi\rangle
 \ee
 and therefore since there is only one generation of oscillators,
 this is symmetric in the indices $\a, \b$, and thus
 $\Phi$ satisfies the wave equation
 \be
 \del_{\adt [\a} \del_{\b] \bdt} \Phi =0 \Longrightarrow \square \Phi = 0.
 \ee
 If we take $\langle \O|= \langle 0|  a_{\a}$ then we obtain a chiral
 spinor field. Again~(\ref{relation}) gives
 \be
 \del_{\a \adt} \Psi_{\b}= \langle 0|a_{\b} a_{\a} b_{\adt}
 \widehat{s}(x) |\Psi\rangle
 \ee
 and this is also symmetric in the indices $\a, \b$ and so
 $\Psi(x)$ must satisfy the spinor equation
 \be
 \del_{\adt [\a} \Psi_{\b]} =0.
 \ee

 \subsection{Unitarity in the bosonic case}

 So far the discussion has simply been about constructing
 representations of the complexified conformal group $SL(4)$. We now
 wish to look at the restriction of these representations to
 representations of the real superconformal group $SU(2,2)$.
 There are two ways of restricting the Lie algebra $\gs \gl(4)$ to $\gs
 \gu(2,2)$. The first choice is known as the compact basis, since in
 this basis the Levi subgroup of $SL(4)$ becomes the compact group $SU(2)\times
 SU(2) \times U(1)$. This is the choice usually associated with the
 oscillator method~\cite{Gunaydin:1998sw,Gunaydin:1998jc,Gunaydin:1999jb}. It is
 obtained by restricting our $\gs \gl(4)$ matrices as
 follows
 \be
 MJ+JM^{\dagger}=0
 \ee
 where $J=\mbox{diag}(1,1,-1,-1)$.
 Note that if $M$ satisfies this condition then from~(\ref{psi})
 \be
 \widehat{M}^{\dagger} +\widehat{M} =0. \la{M+M+}\ee So the representation of
 $SU(2,2)$ obtained by exponentiating $\widehat{M}$ is unitary with
 the natural inner product
 \be
 (\phi,\phi')=\langle \phi |\phi' \rangle.
 \ee

 However, we are interested in the choice of $SU(2,2)$, known as
 the non-compact basis. This is
 defined by the restriction
 \be
  M'K+KM'^{\dagger}=0
 \ee
 where
 \be
 K=\left( \ba{cc} 0_2 & 1_2 \\
                  1_2 & 0_2 \ea \right).
  \ee
 It is this basis for which the induced representations act on fields
 defined on real Minkowski space.

 The two choices for the group $SU(2,2)$ are related by the following
 \be
 M=U^{-1}M'U \la{MU}
 \ee
 where
 \be
 U= {1 \over \sqrt{2}} \left( \ba{cc} 1_2 & 1_2 \\
                                     -1_2 & 1_2 \ea \right)
 \ee
 Since $U$ is an element of the group $SL(4)$, there is a
 corresponding element $\widehat{U}$ in the Fock space of the
 oscillators and we
 obtain the analogue of equation~(\ref{MU})
 \be
  \widehat{M} =\widehat{U}^{-1}\widehat{M'}\widehat{U}.
 \ee
 Then~(\ref{M+M+}) gives
 \be
 \widehat{V} \widehat{M'} + \widehat{M'}^{\dagger} \widehat{V} = 0
 \ee
 where $\widehat{V}=(\widehat{U}\widehat{U}^{\dagger})^{-1}$.
 So in this basis the representations are still unitary, but with an
 unusual inner product
 \be
 (\phi,\phi')=\langle \phi | \widehat{V} |\phi' \rangle.
 \ee

\subsection{The supersymmetric case}

The oscillator construction of the previous section can be
generalised to construct unitary irreducible representations of
the superconformal group
$SU(2,2|N)$~\cite{Bars:1983ep,Gunaydin:1985fk,Gunaydin:1998sw,Gunaydin:1998jc,Gunaydin:1999jb}.
Again we will complexify this and use it to construct the
corresponding irreducible representations of the complex
superconformal group $SL(4|N)$. We will show the relationship
between these and field representations on analytic superspace in
analogy with the above relationship to fields on Minkowski space.
For work on the relationship between oscillator representations
and fields on super Minkowski space in six dimensions
see~\cite{Fernando:2001ak}.

We consider the decomposition of the Lie algebra $\gs \gl (4|N)$
corresponding to $(N, p, N-p)$ analytic space, which has Levi
subgroup $SL(2|p) \oplus SL(2|N-p) \oplus \com$. The Lie algebra
$\gs \gl (4|N)$ consists of matrices $M$ of the following form \be
M= \left(\ba{cc} A^A{}_B & B^{AB'}\\  C_{A'B} & D_{A'}{}^{B'} \ea
\right). \ee where the index $A=(\a,a)$ and $A'=(a',\adt)$ and
$a=1 \dots p, a'=p+1, \dots N$ we can then define generalised
oscillators $\x_A, \x^{A}=\x^{\dagger}_A$ and $\eta_{A'},
\eta^{A'}=\eta^{\dagger}_A$ transforming under the supergroups
$SL(2|p)$ and $SL(2|N-p)$ respectively. The oscillators $\x^a,
\x_a, \eta^{a'},\eta_{a'}$ are odd and all others are even. They
satisfy the following generalised commutation relations \be
[\x_{A}(K),\x^{B}(L)] = \d_{A}^{B} \d_{LK} \qquad
 [ \eta_{A'}(K),\eta^{B'}(L)] = \d^{B'}_{A'}\d_{L K}
 \ee
 where these are commutation relations unless both oscillators are odd
 in which case they are anticommutation relations
 with annihilation (creation) operators labelled by lower (upper)
 indices.

 The Levi decomposition is
 $\gs \gl(4|N)=\gn^-\oplus\gl\oplus\gn^+$ where
 the Levi subgroup corresponds to the block diagonal matrices with
 $B=C=0$, $\gn^-$ and $\gn^+$ correspond to the upper right and lower
 left blocks respectively.
 The `vacuum' is defined by
 \be
 \x_A |0\rangle = \eta_{A'} |0\rangle=0.
 \ee
 We define the objects $\Psi^{\unA}=(\x^A,\eta_{A'})^T$ and
 $\bar{\Psi}_{\unA} =
 \Psi^{\dagger}_{\unB}J^{\unB}_{\unA}=(\x_{\a},-\x_a,-\eta^{a'},-\eta^{\adt})$
 where $J=\mbox{diag}(1_2,-1_p,-1_{N-p},-1_2)$. These satisfy the generalised
 commutation relation
 \be
 [\Psi^{\unA},\bar{\Psi}_{\unB}] = -\d^{\unA}_{\unB}.
 \ee

 For $M \in \gs \gl(4|N)$ we define
 \be
 \widehat M:= -\sum_{K=1}^P
 \bar{\Psi}_{\unA}(K)M^{\unA}{}_{\unB}\Psi^{\unB}(K)
 \ee
 and find that
 \be
 [\widehat M , \widehat M' ] = \widehat{[M,M']}.
 \ee
 So again $\widehat M$ is an infinite dimensional
 representation of
 $\gs \gl(4|N)$ acting on the Fock space of the oscillators.

 The charge $Q$ for $(N,p,N-p)$ analytic superspace (defined as in
 \eq{R(h)}) turns out to be equal to the number of generations of
 oscillators
 \be
 Q=P.
 \ee
 One obtains the irreducible representations by constructing an
 irreducible representation, $\langle\O|,$ of the parabolic subalgebra
 $\gl \oplus \gn^+$ which will in fact be annihilated by $\gn^+$.
 The conjugate representation  $|\O\rangle$ carries an irreducible
 representation of the parabolic subgroup $\gl \oplus
 \gn^-$
 which is annihilated by $\gn^-$. A basis for this representation
 is obtained by acting on $|\O\rangle$ with successive elements of
 $\gn^+$
 \be
 |\O\rangle, \qquad \x^A . \eta^{A'} |\O\rangle, \qquad \x^A . \eta^{A'}
 \x^B . \eta^{B'}|\O\rangle, \dots.
 \ee
 We can then make explicit the relationship between these
 representations and fields on $(N,p,q)$ analytic superspace as in
 the bosonic case. A state $|\Phi\rangle$ corresponds to the
 field $\Phi(X)$ where
 \be
 \Phi(X) = \langle \O | \widehat{s}(X) |\Phi\rangle \label{field}
 \ee
 and where $\widehat{s}(X)=\mbox{exp}(\x_A(K) X^{A A'} \eta_{A'}(K))$
 (with the $K$'s summed over) corresponds to the coset representative
 of $\gs\gl(4|N)$
 for $(N,p,N-p)$ analytic superspace (as in for example~(\ref{s1})).
 Just as in the bosonic case one can check that under the
 $\gs \gl(4|N)$ transformation $|\Phi\rangle \mapsto \widehat{M}
 |\Phi\rangle$ the field defined thus transforms as a field on
 analytic space.\footnote{Note that there is a small difference with the field
 transformations here and the choice defined in~(\ref{tensor}): here the
 combinations $W_A V^A$ and $W^{A'} V_{A'}$ transform as scalars, whereas
 with our previous choice $W^A V_A$ and $W^{A'} V_{A'}$ were scalars.}
 Explicitly
 \be
 |\Phi\rangle = \Phi(0)|\O\rangle + \eta^{A'}\del_{A'A}\xi^{A}\Phi(0)|\O\rangle +
 \dots.
 \ee

For example, if we take $P=1$ then the possible forms $\langle \O |$
can have are
 \be \langle \O |= \langle \O |\x_{A_1} \x_{A_2} \dots \x_{A_s}
 \ee
 and
 \be \langle \O | = \langle \O |\eta_{A'_1} \eta_{A'_2} \dots
 \eta_{A'_{s}}.
 \ee
 So for $(2,1,1)$ analytic superspace consider the case
 $\langle \O |=\langle0|$. According to \eq{field} this
 gives rise to a superfield with no superindices, and a charge $Q=1$,
 i.e. the field $W$ on analytic space corresponding to the
 hypermultiplet~(\ref{hyp}).
 The case $\langle \O | = \langle 0 | \x_{A}$ also
 has one (undashed) superindex and $Q=1$. It therefore gives rise to the field
 $W_A$
 on analytic superspace corresponding to the Maxwell multiplet~(\ref{Max}).

 If we allow two generations of oscillators $P=2$ then we can have
 \be
 \langle\O| = \langle 0| \eta_{A'}(2)\x_A(1)
 \ee
 with charge $Q=2$. This gives the field $T_{A'A}$
 transforming as in~(\ref{stress}) and corresponding to the
 stress-energy multiplet.

\subsection{Unitarity in the supersymmetric case}

 There are two different forms for the real superconformal group
 $SU(2,2|N)$ as there were for the conformal group. The natural form
 for the oscillator construction, and the one taken
 in~\cite{Gunaydin:1998sw,Gunaydin:1998jc,Gunaydin:1999jb} is the
 compact basis where the
 Lie algebra elements $M \in \gs \gl(2,2|N)$ satisfy
 $MJ + JM^{\dagger} = 0$
 where $J=\mbox{diag}(1_2, -1_p, -1_{N-p}, -1_2)$.
 In this case, from the definitions of $\widehat{M}$ and $\bar{\Psi}$
 we find that
 \be
 \widehat{M} + \widehat{M}^{\dagger} = 0 \la{un}
 \ee
 and so the oscillator representations are unitary with the natural
 inner product
 \be
 (\Phi, \Phi')=\langle \Phi | \Phi' \rangle.
 \ee

 We are interested in the non-compact basis of $SU(2,2|N)$, which
 leads to fields on real spacetime. This is defined by restricting $M'
 \in \gs \gl(4)$ to satisfy $M'K + KM'^{\dagger} = 0$ where
 \be
 K= \left(\ba{c|c|c} &&1_2 \\ \hline & 1_N & \\ \hline 1_2 && \ea \right).
 \ee
 These two forms of the Lie algebra $\gs \gl(4|N)$ are related to each
 other by $M=U^{-1}M'U$ where
 \be
 U=\left(\ba{c|c|c}{1 / \sqrt{2}}& &{1 / \sqrt{2}}\\ \hline
                   & 1 & \\ \hline
                   -{1 / \sqrt{2}}&&{1 / \sqrt{2}}\ea \right).
 \ee
 Since $U \in SL(4|N)$ there is a corresponding element
 $\widehat{U}$ in our representation space, and~(\ref{un}) implies that
 \be
 \widehat{V}\widehat{M'}+M'^{\dagger}\widehat{V}=0
 \ee
 where $\widehat{V}=(\widehat{U}\widehat{U}^{\dagger})^{-1}$.
 So representations are unitary with respect to the inner product
 \be
 (\Phi,\Phi')=\langle \Phi | \widehat{V} |\Phi' \rangle.
 \ee

{\bf Acknowledgements}

I am very grateful to my supervisor P.S.Howe for suggestions,
criticism and helpful conversations. This research was supported
by a PPARC studentship and PPARC SPG grant 613.

\appendix
 \section{Appendix}

We here set out some of the general formalism in the purely
bosonic case which we have applied in the supersymmetric case.

\subsection{Roots and parabolic subalgebras}

Let $\gg$ be a complex semisimple Lie algebra and $G$ the
corresponding Lie group. We fix a Cartan subalgebra $\gh$ of $\gg$
and for $\a \in \gh^*$ the dual space of $\gh$ we define
 \be
 \gg_{\a} = \{ v \in \gg : [v,h]=\a(h)v\}.\la{galpha}
 \ee Then the
collection \be \D=\{ \a \in \gh^* : \gg_{\a} \neq {0}, \a \neq 0\}
\ee is called the set of roots of $\gg$ relative to $\gh$. Now
from the Jacobi identity $[\gg_{\a}, \gg_{\b}]\subset \gg_{\a+\b}$
and so the roots span an integral lattice in $\gh^*$. We choose a
system of simple roots of $\gg$ $S=\{\a_i\}$ such that any $\a \in
\D$ can be expressed as a linear combination of elements of $S$
with either all non-negative coefficients, or all non-positive
coefficients. We can choose a different system of simple roots,
but all choices are equivalent up to conjugation ($g\gg g^{-1}$
where $g \in G$ is defined to be a conjugate Lie algebra.) Once we
have defined $S$ we define the positive roots $\D^+$ as \be \D^+ =
\{ \a \in \D : \a = \sum a_i \a_i, \qquad a_i \geq 0, \qquad \a_i
\in S\}. \label{D+} \ee

Now $\gg$ admits a bilinear form known as the Killing form which,
for $u, v \in \gg$, is
\be
 (u,v) =
\mbox{tr } \mbox{ad}(u)\mbox{ad}(v)\label{form} \ee  where ad is
the adjoint representation of $\gg$. This form is non-degenerate
on $\gh$ and we can thus identify $\gh^*$ with $\gh$ by means of
this form: identify $h_{\a} \in \gh$ with $\a \in \gh^*$ where \be
\a(h)=(h,h_{\a}) \forall h \in \gh \label{ha}. \ee We can then
define a bilinear form on $\gh^*$ by \be (\a,\b):=
(h_{\a},h_{\b}). \ee For each root $\a$ we define its co-root
$\a^{\vee}$ as \be \a^{\vee} = 2 \a / (\a,\a). \la{co-root}\ee This always
exists as it can be shown that for all roots $\a$, $(\a,\a)>0$. We
will define $H_i \in \gh$ to be the Cartan subalgebra elements
associated with the simple co-roots
 \be H_i=h_{\a_i^{\vee}}.
 \ee
From this we can define the Cartan Matrix \be c_{ij} = (\a_i,
\a_j^{\vee})=\a_i(H_j) \label{cartan} \ee where $\a_i \in S$ which
uniquely specifies $\gg$. A useful way of describing the Cartan
Matrix is with a Dynkin diagram. This is a graph with nodes
corresponding to the simple roots $\a_i$ and with $L_{jk} =
max{|c_{jk}|,|c_{kj}|}$ edges connecting node $j$ to node $k$.

For the Lie algebra $\gg = \gs \gl (n)$ the Cartan subalgebra is
the set of diagonal matrices (with unit determinant). The roots
are $e_{ij} \in \gh^*$ where $1 \leq i \neq j \leq n$ given by
 \be
 e_{ij}(h) = h_i -h_j\label{roots}
 \ee where $\gh \ni h =
\mbox{diag}( h_1, h_2, \dots h_n)$ and the corresponding subspaces
$\gg_{e_{ij}}$ are spanned by the matrices $\hat{e}_{ji} \in \gs \gl(N)$
with zeros everywhere except
in the $ji$ component where there is a 1. Simple roots can be chosen
as
 \be \a_i=
 \{e_{i(i+1)}: i=1 \dots n-1\}. \ee
In this case the bilinear form~(\ref{form}) simplifies to
 \be
(u,v) = 2N \mbox{ tr} (uv).
  \ee
 The simple co-roots are $\a_i^{\vee}=2N \a_i$ and the
 corresponding member of the Cartan subalgebra
 is
 \be
 H_i=\hat{e}_{ii}-\hat{e}_{(i+1)(i+1)}.\label{hv}
 \ee
One can then work out the Cartan matrix and the Dynkin diagram.
The Dynkin diagram for $\gs\gl(N)$ is given by $N-1$ nodes each
connected by single edges to the adjacent nodes
 \be
 \begin{picture}(120,20)(-10,-10)
\put(0,0){\makebox[0pt][l]{$\underbrace{\bt \hspace{1.5em}
\bt\hspace{1.5em}\cdots
\hspace{1.5em}\bt\hspace{1.5em}\bt}_{N-1}$}
\rule[.5ex]{2.2em}{.1ex} $\hspace{4.3em}$
\rule[.5ex]{2.2em}{.1ex}}
\end {picture}
\ee

The standard Borel subalgebra $\gb$ is given by \be \gb=\gh \oplus
\gn \label{borel} \ee where \be \gn = \bigoplus_{\a \in \D^+}
\gg_\a. \label{nil} \ee A parabolic subalgebra $\gp$ is defined to
be one which contains the Borel subalgebra. Alternatively specify
a subset $S_{\gp}$ of $S$ and define \be \D(l):=\mbox{Span
}S_{\gp} \cap \D \label{D} \ee where Span $S_{\gp}$ is the
subspace of $\gh^*$ spanned by linear combinations of the elements
of the root vectors in $S_{\gp}$. Then define \be \gl:= \gh
\oplus\bigoplus_{\a \in \D(\gl)} \gg_\a \ee and \be \gn =
\bigoplus_{\a \in \D(\gn)}\gg_{\a} \ee where $\D(\gn)=
\D^+\bsh\D(\gl).$ The subalgebra defined as \be \gp := \gl \oplus
\gn. \label{p} \ee contains the Borel subalgebra and is thus
parabolic, and up to conjugation, all parabolics arise in this way
(i.e. all parabolics are of the form $g^{-1} \gp g$ for some $g
\in G$ and some standard parabolic $\gp$ constructed as above.)
The subgroup $\gl$ is known as the Levi subgroup.

For $\gs \gl (N)$ the standard Borel subalgebra is given by the
set of traceless lower triangular matrices, and the standard
parabolic subalgebras are the set of traceless `block
lower-triangular' matrices:
\be
 \begin{picture}(300,100)(-30,-40)
 \put(10,0){$\left(\ba{cccccccccc}
  \hspace{.2em} \bt \hspace{.2em}&\hspace{.2em} \bt \hspace{.2em}& &&& &&&&    \\
  \hspace{.2em} \bt \hspace{.2em}&\hspace{.2em} \bt \hspace{.2em}&&&&  &&&&\\
  \hspace{.2em} \bt \hspace{.2em}&\hspace{.2em} \bt \hspace{.2em}&\hspace{.2em} \bt \hspace{.2em}&\hspace{.2em} \bt \hspace{.2em}& & &&&&\\
  \hspace{.2em} \bt \hspace{.2em}&\hspace{.2em} \bt \hspace{.2em}&\hspace{.2em} \bt \hspace{.2em}&\hspace{.2em} \bt \hspace{.2em}& & &&&&\\
  \hspace{.2em} \bt \hspace{.2em}&\hspace{.2em} \bt \hspace{.2em}&\hspace{.2em} \bt \hspace{.2em}&\hspace{.2em} \bt \hspace{.2em}&\hspace{.2em} \bt \hspace{.2em}&\hspace{.2em} \bt \hspace{.2em}&\hspace{.2em} \bt \hspace{.2em}& &&\\
  \hspace{.2em} \bt \hspace{.2em}&\hspace{.2em} \bt \hspace{.2em}&\hspace{.2em} \bt \hspace{.2em}&\hspace{.2em} \bt \hspace{.2em}&\hspace{.2em} \bt \hspace{.2em}&\hspace{.2em} \bt \hspace{.2em}&\hspace{.2em} \bt \hspace{.2em}& &&\\
  \hspace{.2em} \bt \hspace{.2em}&\hspace{.2em} \bt \hspace{.2em}&\hspace{.2em} \bt \hspace{.2em}&\hspace{.2em} \bt \hspace{.2em}&\hspace{.2em} \bt \hspace{.2em}&\hspace{.2em} \bt \hspace{.2em}&\hspace{.2em} \bt \hspace{.2em}& &&\\
  &&&&&&&& .& \\
  &&&&&&&&&.
  \ea\right)$}
  \put(25,45) {$
 \left.\phantom{\ba{cc}
  \bullet&\bullet\\
  \bullet&\bullet\ea}\right\}
{\scriptscriptstyle k_1} $}
  \put(30,35) {$
 \left.\phantom{\ba{cccc}
  \bullet&\bullet&\bt&\bt\\
  \bullet&\bullet&\bt&\bt\\
  \bullet&\bullet&\bt&\bt\\
  \bullet&\bullet&\bt&\bt
 \ea}\right\}
{\scriptscriptstyle k_2} $}
  \put(40,15) {$
 \left.\phantom{\ba{ccccccc}
  \bullet&\bullet&\bt&\bt&\bt&\bt&\bt\\
 \bullet&\bullet&\bt&\bt&\bt&\bt&\bt\\
 \bullet&\bullet&\bt&\bt&\bt&\bt&\bt\\
 \bullet&\bullet&\bt&\bt&\bt&\bt&\bt\\
 \bullet&\bullet&\bt&\bt&\bt&\bt&\bt\\
 \bullet&\bullet&\bt&\bt&\bt&\bt&\bt\\
 \bullet&\bullet&\bt&\bt&\bt&\bt&\bt\ea}
  \right\}{\scriptscriptstyle k_3} $}
\end{picture}\label{para}
 \ee
 The corresponding Levi subalgebra is the block diagonal subalgebra,
i.e. $\gs(\gg \gl (k_1) \oplus \gg \gl (k_2-k_1) \oplus \ldots)$.

A useful notation for a standard parabolic $\gp \subseteq \gg$
is to cross through all the nodes in the Dynkin diagram
corresponding to the simple roots in $S\bsh S_{\gp}$. So the
parabolic in our example~(\ref{para}) would be represented by
placing a cross on each of the nodes $k_1,k_2, \ldots, k_l$ of the
Dynkin diagram for $\gs\gl(N)$. The Borel subalgebra is
represented by a Dynkin diagram with all nodes crossed out.

\subsection{Representations}

If $W$ is a representation space for $\gg$ then $0 \neq w \in W$
is a weight vector of weight $\l \in \gh^*$ if and only if
 \be
 R(h) w = \l(h) w \qquad \forall h \in \gh \label{weights}
 \ee where
$R$ is the representation. We define raising and lowering
subalgebras of $\gg$ respectively as \be \gn= \bigoplus_{\a \in
\D^+} \gg_{\a}; \qquad \gn_-= \bigoplus_{\a \in \D^-} \gg_{\a}.
\ee (For $\gs \gl (N)$, $\gn$ is the set of lower triangular
matrices excluding the diagonals, and $\gn_-$ is the set of upper
triangular matrices excluding the diagonals.)
 Note that if $w$ is a weight vector of weight $\l$, and $v \in \gg_{\a}$
 then~(\ref{weights},\ref{galpha})) give
 \be
 R(h)R(v)w=(\l-\a)(h)R(v)w
 \ee
 and so $R(v)w$ is a weight vector with weight $\l-\a$. We
 define a partial ordering on the root space by
 \be
 \l \preceq \l' \Leftrightarrow \l=\l' + a_i \a_i, \qquad a_i \geq 0,
 \qquad \a_i \in \D^+. \la{po}
 \ee
 So applying lowering operators on a weight space of weight $\l$
 produces a weight
 space with weight $\l'\preceq \l$. A maximal weight state
 $w\in W$ with weight $\l$ is
 annihilated by all raising operators and a maximal weight state is a
 highest weight state if $\l \succeq \l' \ \forall \l' \in \D$. A set of weights $\l_i$
 dual to the simple roots are
defined by \be (\l_i, \a_j^{\vee})=\d_{ij}. \label{li} \ee Then
any weight $\l$ is expressed uniquely as \be \l=\sum_i
(\l,\a^{\vee}_i)\l_i.\ee We can then represent a weight $\l$ by
putting the coefficient
 \be
 n_j=(\l,\a_j^{\vee})=\l(H_j)\label{nodes}
 \ee
 over
the node of the Dynkin diagram for $\gg$ which corresponds to the
root $\a_j$. A weight $\l$ is said to be integral if all the
coefficients are integers and dominant if they are all
non-negative.

Now it turns out that any finite dimensional representation of
$\gg$ has a unique highest weight vector which is dominant
integral and there is a one-to-one correspondence between
dominant integral highest weight states and finite dimensional
irreducible representations. We will however be mostly interested
in infinite dimensional representations, which still have a unique
highest weight state. These will be called highest weight
representations. We will thus use the Dynkin diagram with Dynkin
coefficients, both to describe the weight, and also the
irreducible representation which has this weight as its highest
weight.

A similar statement can be made for representations of a parabolic
subalgebra $\gp$. In fact, if $\gp = \gl \oplus \gn$ is a Levi
decomposition of $\gp$ then $\gn$ acts trivially on any
irreducible representation of $\gp$ and so an irreducible
representation of $\gp$ corresponds to an irreducible
representation of the Levi subalgebra $\gl$. $\gl$ splits into a
semisimple part and a centre \be \gl = \gl_S +\gl_Z \mbox{ where }
\gl_S = [\gl,\gl]. \ee So an irreducible representation of $\gp$
corresponds to an irreducible representation of $\gl_S$ together
with $\com$-charges giving a representation of $\gl_Z$ and we know
that the representation of $\gl_S$ has a unique highest weight
from above. A weight $\l \in \gh^*$ is said to be integral or
dominant for $\gp$ if $\t(\l)$ is dominant or integral for $\gl_S$
where $\t:\gh^* \rightarrow (\gh \cap \gl)^*$ is the natural
projection operator. It is then clear that finite dimensional
representations of $\gp$ are in one-to-one correspondence with
weights which are dominant integral for $\gp$. Diagrammatically
this is represented by insisting that all the coefficients
corresponding to nodes which are not crossed through in the Dynkin
diagram must be non-negative integers, i.e. $(\l,\a_i^{\vee})\in
\bbZ^+ \cup\{0\}$ for all $\a_i \in S_{\gp}$. Furthermore, in
order for the representation to exponentiate to a representation
of the Lie group $P$, the highest weight must be integral (but not
necessarily dominant) for $\gg$ and not just for $\gp$. In other
words finite dimensional irreducible representations of $P$
correspond to Dynkin diagrams with crosses through some nodes and
integral coefficients above the nodes, where the integers above
nodes which are not crossed out must be non-negative.

\subsection{Tensor representations}

All finite dimensional irreducible representations of $\gs \gl(n)$
can be given explicitly as tensor products of the fundamental
representation. These tensor products are symmetrised to put them
into irreducible representations. The symmetry properties of these
tensor products is given using Young Tableaux in the usual way.
 \be
\setlength{\unitlength}{.3mm}
\begin{picture}(420,120)(0,40)
\put(20,140){\framebox(400,20){}}\put(30,146){$m_1$}
\put(20,120){\framebox(340,20){}}\put(30,126){$m_2$}
\put(20,100){\framebox(280,20){}}
 \put(20,80){\framebox(220,20){}}
\put(20,60){\framebox(160,20){}}\put(30,66){$m_{n-2}$}
\put(20,40){\framebox(100,20){}}\put(30,46){$m_{n-1}$}
\end{picture}
\la{tab} \ee
 where there are $m_i$ squares in each row. To convert from a
 Young Tableau notation to a Dynkin diagram for a particular
 representation one we must find the highest weight state of the
 tensor representation. For $\gs\gl(n)$ this is
 obtained by putting 1's in the first row of the Young Tableau,
 2's in the second row, and so on. From
~(\ref{weights},\ref{nodes}) we see that
 \be
 R(H_i)w=n_i w
 \ee
 if $w$ is the highest weight state. Applying this formula and our
 expression for $H_i$~(\ref{hv}) we obtain
 \be
 n_i = m_i -m_{i+1}.
 \ee

\subsection{Induced representations}

One way of giving representations of a group is with induced
representations which we now describe. If $G$ is a Lie group and
$P$ is a subgroup of $G$ we form the coset space $X$ of right
cosets $X=P\bsh G$. Fields on these spaces are equivariant maps
$F:G\rightarrow V$, where $V$ is a representation space for $P$,
i.e. maps such that
\be F(hu)=R(h)F(u) \label{F(hu)=}
\ee
where
$u\in G$, $h \in P$
 and $R$ is the representation of $P$ on $V$. These fields form
 representations of $G$ as follows:
\be (g\cdot F)(u)=F(ug).
 \ee In practice one uses local sections
$s:U \rightarrow G$ of the bundle $G \rightarrow H\bsh G$, $U\ni x
\mapsto s(x)$ where $U$ is an open subset of $X$. Under a group
transformation $g \in G$, $x \mapsto x'$ where $x'$ can be found
by the following formula: \be s(x)g=h(x,g)s(x') \ee $h\in P$. Thus
 \be
 (g.F)(s(x))=R(h(x,g))F(s(x')).\label{g.F}
 \ee

The method of parabolic induction involves induced representations
of a complex, simple Lie group $G$ with a standard parabolic
subgroup $P$, the maps $F$ being holomorphic on $P\bsh G$. In this
case the coset space $X=P \bsh G$ is known as a flag manifold. For
example complexified, compactified Minkowski space can be viewed
as the coset space $P\bsh SL(4)$, $SL(4)$
 being the complexified conformal group and $P$ the parabolic subgroup of
 matrices of the following shape:
 \be
 \left(\ba{cccc} \hspace{.2em} \bt \hspace{.2em}&\hspace{.2em} \bt
  \hspace{.2em}&&\\ \hspace{.2em} \bt \hspace{.2em}&\hspace{.2em}
  \bt \hspace{.2em}&&\\
 \hspace{.2em} \bt \hspace{.2em}&\hspace{.2em} \bt \hspace{.2em}
 &\hspace{.2em} \bt \hspace{.2em}&\hspace{.2em} \bt \hspace{.2em}\\
 \hspace{.2em} \bt \hspace{.2em}&\hspace{.2em} \bt \hspace{.2em}
 &\hspace{.2em} \bt \hspace{.2em}&\hspace{.2em} \bt \hspace{.2em}
 \ea
 \right)\label{shape}
\ee
where the bullets denote elements which do not have to be
zero. The blank region can be thought of as corresponding to
spacetime. Indeed, it takes six charts to cover the whole of
compactified complexified Minkowski space. We focus on the coset representative of the
form
\be X\ni x\mapsto s(x)= \left(\ba{cc} 1_2&x\\ 0_2&1_2 \ea
\right)\label{coset}
\ee
 which can be identified with non-compact (affine) Minkowski
 space. Here each entry is a two-by-two matrix and $x$ is a matrix
 representation of the coordinates for
 complexified Minkowski space. The transformation of $x$ under the
 conformal group can easily be calculated using the above
 formalism, and agrees with the usual one. Here the Levi subalgebra is
 $\gs(\gg \gl
(2) \oplus \gg \gl(2))$ and the Dynkin diagram is
\begin{picture}(30,10)(0,0) \put(0,0){\makebox[0pt][l]{$\bt
 \xz\bt$} \rule[.5ex]{1.8em}{.1ex} } \end {picture}.

To specify an induced representation on a flag manifold $P \bsh G$
we need to give the representation $R$ of the parabolic subalgebra
$P$ under which our fields transform. But the finite dimensional
irreducible representations of $P$ have been described above. They
can be represented by a Dynkin diagram with crosses through some
nodes and integer coefficients above each node, the integers above
uncrossed nodes being non-negative. We use this same Dynkin
diagram to describe the induced representation.

The Borel-Weil theorem states that if $\l \in \gh^*$ is dominant
integral for $G$ then for any parabolic subgroup $P$ the induced
representation specified by this $\l$ is isomorphic to the unique
finite dimensional irreducible representation specified by the
highest weight $\l$. So diagrammatically, if $a_i \geq 0$ for all
$i$ then
 \be
\ba{rcl} &&\begin{picture}(150,20)
\put(0,0){\makebox[0pt][l]{$\bt\hspace{1.5em}\bt
\hspace{1.5em}\bt\hspace{1.5em} \bt \hspace{1.5em}\bt
\hspace{1.5em}\bt\hspace{1.5em}\bt$} \rule[.5ex]{2em}{.1ex}
$\hspace{0.3em} \cdots \hspace{.9em} \cdots \hspace{.9em} \cdots
\hspace{0.9em} \cdots \hspace{0.3em} $ \rule[.5ex]{2em}{.1ex} }
\put(0,10){\tiny $ a_1 \hspace{2em} a_2\hspace{2em} a_{j}
\hspace{2em} a_k \hspace{2em} a_{l}\hspace{2em}
a_{N-2}\hspace{1em} a_{N-1}$}
\end {picture}\\
&\cong&
\begin{picture}(150,30)
\put(0,0){\makebox[0pt][l]{$\bt\hspace{1.5em}\bt
\hspace{1.5em}\times
  \hspace{1.5em}
\bt \hspace{1.5em}\times \hspace{1.5em}\bt\hspace{1.5em}\bt$}
\rule[.5ex]{2em}{.1ex} $\hspace{0.3em} \cdots \hspace{1.2em}
\cdots \hspace{1em} \cdots \hspace{1em} \cdots \hspace{0.3em} $
\rule[.5ex]{2em}{.1ex} } \put(0,10){\tiny  $a_1 \hspace{2em}
a_2\hspace{2.5em} a_{j} \hspace{2.5em} a_k \hspace{2em}
a_{l}\hspace{2em} a_{N-2}\hspace{1em} a_{N-1}$}
\end {picture}
\ea \ee where crosses can be placed on any of the nodes of the
right hand side. Furthermore, if $\l$ is not dominant integral for
$G$ then the corresponding induced representation is 0.

\subsection{Unitary irreducible representations of the conformal group}

Representations of the conformal group $SU(2,2)$ are usually
specified by the quantum numbers $J_1,J_2$ the spins, and $L$ the
dilation weight. These are related to the three Dynkin
coefficients $n_i$ as follows:
 \be
 n_1 = 2J_1 \qquad n_2 = -L - J_1 - J_2 \qquad n_3 = 2J_2. \label{dy}
 \ee
 Unitary irreducible highest
weight representations satisfy the following unitary bounds
\cite{Mack:1977je} which we shall call a), b) and c) to tie in with the
supersymmetric case:
 \be \ba{crrclcrrcl}
  \mbox{a)}& J_1 J_2 \neq 0,& L&\geq& J_1 + J_2 + 2 &\Rightarrow& n_1 n_3 \neq0,& n_1+n_2+n_3 & \leq&-2\\
  \mbox{b)}& J_1 J_2 = 0,& L&\geq& J_1 + J_2 + 1 &\Rightarrow& n_1 n_3=0,& n_1+n_2+n_3 & \leq&-1\\
  \mbox{c)}& J_1 =J_2= 0,& L&=& 0 &\Rightarrow& n_1 =n_3 =0,& n_2 & =&0\\
  \ea \la{bbounds}
  \ee
 Note that the central node $n_2$ is negative for all non-trivial
 representations, and thus according to the Borel-Weil theorem the
 representations are 0 if given as fields on complexified compactified
 Minkowski space $P\bsh SL(4)$ where P is given in~(\ref{shape}). However
 we will be considering fields on affine
 Minkowski space which is an open subset of $P\bsh G$ and thus the
 Borel-Weil theorem no longer holds. This is because we are really
 considering representations of the non-compact group $SU(2,2)$.
 The negative node means
 that these representations must be infinite dimensional
 representations as all finite dimensional representations have
 dominant integral highest weights. The fact that we do not consider
 the whole of the space $P\bsh SL(4)$ also means that the induced
 representations are not always irreducible (see section \ref{sect:red}).

\subsection{Projective twistor space}\label{twistor}

It will be instructive to consider the twistor formalism, as we
will be using similar techniques, although things are much simpler
in the supersymmetric case. In twistor theory one considers a
double fibration from complex Minkowski space. Projective Twistor
space, $\bbP \bbT$, is a subspace of $\com \bbP^3$ which is the
flag space $P\bsh SL(4)$, given by the following Dynkin diagram
 \be
 \com \bbP^3= \begin{picture}(30,10)(0,0) \put(0,0){\makebox[0pt][l]{$\xz
\hspace {2em}
 \bt \hspace{2em}\bt$} \rule[.5ex]{5.5em}{.1ex} }
 \end{picture}
\ee
 There is a double fibration between $\com \bbP^3$ and complex
 compactified Minkowski space which can be given in terms of Dynkin
 diagrams as
 \begin{equation}
\begin{picture}(300,120)(-80,-10)
\put(53,80){\makebox[0pt][l]{$\xz \hspace {1em}
 \xz \hspace{1em}\bt$} \rule[.5ex]{4em}{.1ex}} \put(60,68){\vector(-1,-1){50}}
\put(80,68){\vector(1,-1){50}} \put(-10,0){\makebox[0pt][l]{$\xz
\hspace {1em}
 \bt \hspace{1em}\bt$} \rule[.5ex]{3.7em}{.1ex}}
\put(60,0){$\Longleftrightarrow$}
\put(130,0){\makebox[0pt][l]{$\bt \hspace {1em}
 \xz \hspace{1em}\bt$} \rule[.5ex]{3.7em}{.1ex}}
\put(11,50){$\pi_L$} \put(111,50){$\pi_R$}
\end{picture}
\end{equation}
 This induces a sub-double fibration from complex affine
 Minkowski space, $\bbM$ as follows
 \begin{equation}
\begin{picture}(300,120)(-80,-10)
\put(70,80){$\bbF$} \put(60,68){\vector(-1,-1){50}}
\put(80,68){\vector(1,-1){50}} \put(-10,0){${\bbP \bbT}$}
\put(60,0){$\Longleftrightarrow$} \put(130,0){${\bbM}$}
\put(11,50){$\pi_L$} \put(111,50){$\pi_R$}
\end{picture}
\end{equation}
 so that $\bbF=\pi^{-1}_R \bbM$ and ${\bbP \bbT}=\pi_L\bbF$. One
 then wishes to look at unitary irreducible representations of the
 conformal group. If we considered these representations as holomorphic
 fields on twistor space we would have the following Dynkin diagram
 \be
 \begin{picture}(30,10)(0,0) \put(0,0){\makebox[0pt][l]{$\xz
\hspace {2em}
 \bt \hspace{2em}\bt$} \rule[.5ex]{5.5em}{.1ex} }
 \put(0,10){\tiny $n_1$ \hspace{2.5em} $n_2$ \hspace {2em} $n_3$}
 \end{picture}
\ee
 and since $n_2 < 0$ we do not have a finite dimensional
 representation of the parabolic subgroup. In twistor theory one
 gets around this by looking at higher cohomology. The result is
 that one can obtain the irreducible highest weight
 representations as higher cohomology classes on Twistor spaces
 without enforcing any constraints. In the supersymmetric case we
 do something similar, but
 things are much simpler as we do not have any negative nodes and
 therefore one does not have to consider higher cohomology
 classes. By looking at suitable spaces, one can obtain all
 unitary irreducible representations of the superconformal group
 as unconstrained holomorphic fields on these spaces.

\providecommand{\href}[2]{#2}\begingroup\raggedright\endgroup

\end{document}